\documentclass[aps,amsmath,amssymb,prd,showpacs,floatfix,preprint,superscriptaddress,nofootinbib,12pt]{article}
\usepackage{jheppub}
\usepackage{slashed}
\pdfoutput=1
\usepackage{fancyhdr}
\makeatletter
\def\@fpheader{\relax}
\makeatother

\usepackage[svgnames]{xcolor}
\usepackage[framemethod=tikz]{mdframed}
\usepackage[customcolors]{hf-tikz}
\usetikzlibrary{shadows}

\usepackage{amsmath,epsfig}
\usepackage{amssymb,amsfonts}
\usepackage{latexsym}
\usepackage{subeqnarray}
\usepackage{xcolor}

\usepackage{graphicx}
\usepackage{ulem}

\usepackage[utf8]{inputenc}

\relax
\def\be{\begin{equation}}
\def\ee{\end{equation}}
\def\bea{\begin{eqnarray}}
\def\eea{\end{eqnarray}}

\newcommand\fverb{\setbox\pippobox=\hbox\bgroup\verb}
\newcommand\fverbdo{\egroup\medskip\noindent%
                        \fbox{\unhbox\pippobox}\ }
\newcommand\fverbit{\egroup\item[\fbox{\unhbox\pippobox}]}

\newcommand{\bear}{\begin{eqnarray}}

\newcommand{\eear}{\end{eqnarray}}

\newcommand{\bsea}{\begin{subeqnarray}}
\newcommand{\esea}{\end{subeqnarray}}
\newbox\pippobox

\def\6{\partial}

\newcommand{\comments}[1]{}
\input Starburst.fd

\allowdisplaybreaks[3]

 \title{\center \Huge Low frequency propagating shear waves
in holographic liquids}

\author[\,\,\clubsuit,\,\spadesuit]{Matteo Baggioli}
\affiliation[\clubsuit]{Instituto de Fisica Teorica UAM/CSIC, c/Nicolas Cabrera 13-15,
Universidad Autonoma de Madrid, Cantoblanco, 28049 Madrid, Spain.}
\affiliation[\spadesuit]{Crete Center for Theoretical Physics, Institute for Theoretical and Computational Physics\\ Department of Physics, University of Crete, 71003
Heraklion, Greece.}
\author[\bigstar]{\& Kostya Trachenko}
\affiliation[\bigstar]{ School of Physics and Astronomy, Queen Mary University of London, Mile End Road, London, E1 4NS, UK}

\emailAdd{matteo.baggioli@uam.es}
\emailAdd{k.trachenko@qmul.ac.uk}

\abstract{Recently, it has been realized that liquids are able to support solid-like transverse modes with an interesting gap in momentum space developing in the dispersion relation. We show that this gap is also present in simple holographic bottom-up models, and it is strikingly similar to the gap in liquids in several respects. Firstly, the appropriately defined relaxation time in the holographic models decreases with temperature in the same way. More importantly, the holographic $k$-gap increases with temperature and with the inverse of the relaxation time. Our results suggest that the Maxwell-Frenkel approach to liquids, involving the additivity of liquid hydrodynamic and solid-like elastic responses, can be applicable to a much wider class of physical systems and effects than thought previously, including relativistic models and strongly-coupled quantum field theories. More precisely, the dispersion relation of the propagating shear waves is in perfect agreement with the Maxwell-Frenkel approach. On the contrary the relaxation time appearing in the holographic models considered does not match the Maxwell prediction in terms of the shear viscosity and the instantaneous elastic modulus but it shares the same temperature dependence.}

\preprint{CCTP-2018-8, ITCP-IPP 2018/6}
\begin{document}
\maketitle
\fancyhf{}
\renewcommand*{\thefootnote}{\arabic{footnote}}	
\setcounter{footnote}{0}
\section{Introduction}
The ability to flow generally differentiates liquids from solids. Making this difference quantitative on the basis of a physical theory has proved to be an important open question. More recently, it has been appreciated that similarities between liquids and solids involve an important length scale.
Liquids are traditionally discussed in the hydrodynamic regime $\omega/T \ll 1, k/T \ll 1$ where no shear waves propagate \cite{Landau1987Fluid}. From a theoretical point of view, the main traditional distinction between liquids and solids has been their ability to support propagating shear waves: solids do support propagating transverse phonons while liquids do not\footnote{For the rest of the manuscript we will refer to this point as the distinction between solids and liquids. Notice that such a difference relies on very specific transverse excitations, \textit{i.e.} transverse sound. It is clear that also in liquids there could be propagating transverse waves of other kind, such as EM waves.}. This property is connected to the existence of a finite elastic shear modulus which gives solids the ``rigidity'' properties we are used to. Indeed, it is very well-known \cite{landau7,Lubensky} that transverse phonons in solids obey the simple dispersion relation:
\begin{equation}
\omega\,=\,\mathcal{V}\,k\,,\quad \text{with}\quad \mathcal{V}^2\,=\,\frac{G}{\chi_{PP}}
\end{equation}
where $G$ is the shear modulus, $\chi_{PP}$ the momentum susceptibility and $\mathcal{V}$ the speed of shear sound. \\

However, this picture is not the complete story and was challenged by Frenkel long time ago, in 1932 \cite{doi:10.1021/j150454a025}. Accordingly to his ideas, above a certain frequency and away from the hydrodynamic regime where:
\begin{equation}
\omega\,>\,\omega_F\,\equiv \frac{1}{\tau}\label{we}
\end{equation}
liquids, in fact, can support propagating shear waves and therefore behave like solids. Here, $\tau$ is liquid relaxation time introduced by Frenkel, the time between consecutive particle jumps in the liquid. Frenkel’s argument was that at times shorter than $\tau$, the system is a solid and therefore supports all three phonon modes, one longitudinal and two transverse. \cite{doi:10.1021/j150454a025}. At times longer than $\tau$, the system is hydrodynamic and supports one longitudinal mode only.

Interestingly, Frenkel's proposal implied that the difference between a liquid and an isotropic solid (glass) is only quantitative (because $\tau$ grows continuously with reducing temperature) but not {\it qualitative}, sparking an interesting debate between him and Landau (see Ref. \cite{2016RPPh...79a6502T}) for details). The predictions of Frenkel theory have been confirmed in several simulations and experiments \cite{2016RPPh...79a6502T}.\\
Very recently, more challenges from rheology experiments appeared. In particular, new results seem to suggest that liquids can support propagating shear waves and behave like solids even at low frequencies \cite{Mendil2006,doi:10.1002/pi.2625,NOIREZ201016,noirez2012identification}.  This behaviour does not seem to be consistent with the Frenkel's proposal although, as we will see below, Frenkel did have another equation in his book capable of explaining the low-frequency shear resistance of liquids. 

Two proposals have been put forward to explain the low-frequency behavior of liquids:
\begin{enumerate}
\item The existence of a k-gap in the transverse spectrum of liquids \cite{PhysRevLett.118.215502,2016RPPh...79a6502T,PhysRevE.96.062134,2013NatSR...3E2794B}
\item The role of non-affine deformations in  viscoelastic materials and amorphous solids \cite{PhysRevB.83.184205,PhysRevLett.110.178002,PhysRevB.90.140203}
\end{enumerate}
According to the first proposal, transverse shear waves in a liquid have dispersion relation of the type:
\begin{equation}
Re(\omega)\,=\,\sqrt{\mathcal{V}^2\,k^2\,-\,\frac{1}{4\,\tau^2}}\label{ddd}
\end{equation}
As discussed below, this result can be derived from the Navier-Stokes equation using the so-called \textit{Maxwell interpolation} \cite{2016RPPh...79a6502T}. This result implies that at large enough momenta
\begin{equation}
k\,>\,k_{gap}\,\equiv \frac{1}{2\,\mathcal{V}\,\tau}
\end{equation}
liquids can support propagating shear waves \underline{even} at low frequency $\omega \approx 0$. In other words, liquids could behave like solids even at very low frequencies. This observation may provide an explanation for the unusual experimental results obtained in \cite{Mendil2006,doi:10.1002/pi.2625,NOIREZ201016,noirez2012identification}. We will analyze in more details the theoretical background of both Frenkel ideas and the k-gap proposal in section \ref{prologue}. To avoid any confusion it is important to clarify that the propagating shear waves at large frequency or large momentum cannot be interpreted in terms of phonons as Goldstone bosons for translational invariance and cannot be captured by a simple low energy effective field theory. That said, for simplicity, we will use interchangeably the terms ''propagating shear waves'' and ''propagating transverse phonons' in the rest of the manuscript.\\

In a very different area of physics, the holographic correspondence has been very useful in the understanding hydrodynamics of strongly coupled fluids\footnote{In this context the theories at hand have usually a relativistic UV fixed point. For that reason the asymptotic speed appearing in the $k$-gap dispersion relation \eqref{ddd} will be simply the speed of light $\mathcal{V}=c$.}. Several important results have been obtained using such a technique \cite{Policastro:2001yc,Landsteiner:2011cp}. Moreover, the collective excitations of the dual field theory are easily addressed in the AdS-CFT framework. These excitations are simply encoded in the QNM frequencies of specific gravitational black hole solution \cite{Son:2002sd,Policastro:2002se,Berti:2009kk}. Finally, in the last years, holography has been extended to the identification of gravity duals for solids and viscoelastic materials which exhibit elastic properties and propagating transverse shear waves \cite{Baggioli:2014roa,Baggioli:2015gsa,Alberte:2015isw,Alberte:2016xja,Baggioli:2016rdj,Andrade:2017cnc,Alberte:2017cch,Alberte:2017oqx,Baggioli:2018bfa,Esposito:2017qpj,Amoretti:2017frz,Grozdanov:2018ewh}.\\[0.2cm]
In this work we will ask the following simple questions:
\begin{center}
\textit{How are solid-like properties of liquids encoded in holographic models?}\\[0.05cm]
\textit{If yes, what is the associated time scale $\tau$?}
\end{center}
As we discuss below, the answer is positive to the first two questions, and moreover we  will be able to identify precisely the corresponding relaxation time in the problem.\\

In this paper, we analyze two different holographic models \cite{Andrade:2013gsa},\cite{Grozdanov:2018ewh} which, similarly to liquids, exhibit the presence of a $k$-gap in the spectrum of transverse collective modes. Importantly and unexpectedly, the temperature dependence of the gap is the same as in the condensed matter system, liquids. \\
The discussed features go beyond the usual hydrodynamic limit\footnote{By hydrodynamic we mean the limit of small frequencies and momenta compared to the thermal energy scale:
\begin{equation}
\omega\,\ll\,T\,,\quad k\,\ll\,T\,.
\end{equation}} and need a detailed numerical study at large momenta $k\gg T$. Despite this difficulty we obtain an analytic formula for the liquid relaxation time which can  be defined just in terms of hydrodynamic quantities and it is in very good agreement with our numerical data. \textit{En passant} we study the thermodynamic properties of the holographic models and we emphasize their similarities with the behaviour of amorphous solids, glasses and viscoelastic materials.\\[0.1cm]
We believe this work can constructively contribute to answering our initial question about physical differences between solids and liquids.
Additionally, our results provide a new unexpected and extraordinary link between condensed and soft matter physics on one hand, and black hole physics on the other.\\[0.2cm]
The paper is organized as follows: in section \ref{prologue}, we discuss the theoretical problems of the liquid description, Maxwell-Frenkel approach to liquids involving the additivity of hydrodynamic and solid-like elastic response and show how this perspective results in the emergence of the gap in the liquid transverse spectrum in $k$-space; in sections \ref{hol22}, \ref{hol1} we present two simple existing holographic models exhibiting a k-gap in accordance to the Condensed Matter theories; finally in section \ref{disc} we conclude discussing the results and some possible future questions.
\section{Condensed matter prologue}\label{prologue}

A theory of weakly-interacting gases can be developed using perturbation theory. On the other hand, interactions in a liquid are strong. Therefore, the liquid energy and other properties are strongly system-dependent. For this reason, a theory of liquids was believed to be impossible to construct in general form \cite{landau}.

Perturbation theories do not apply to liquids because inter-atomic interactions are strong as already noted. Solid-based approaches seemingly do not apply to liquids either: its unclear how to apply the traditional harmonic expansion around equilibrium positions because the equilibrium lattice does not exist to begin with, due to particle re-arrangements that enable liquids to flow. This combination of strong interactions and large particle displacements has proved to be the ultimate problem in understanding liquids theoretically, and is known as the ''absence of a small parameter''.

The absence of traditional simplifying features in the liquid description does not mean that the problem can not be solved in some other way, including attempting to solve the problem from first-principles using the equations of motion. Interestingly, this involves solving a large number of non-linear equations which is exponentially complex and therefore is not currently tractable \cite{2016RPPh...79a6502T}.

Recent progress in understanding liquids followed from considering what kind of collective modes (phonons) can propagate in liquids and supercritical fluids \cite{2016RPPh...79a6502T}. It has been ascertained that, (a) the $k$-gap $k_g$ increases with the inverse of liquid relaxation time in a wide range of temperature and pressure for different liquids and supercritical fluids and (b) $k_g$ increases with temperature. The first result directly agrees with \cite{PhysRevLett.118.215502}; the second result agrees with \cite{PhysRevLett.118.215502} noting that $\tau$ decreases with temperature.

Below we briefly review the origin of this result. This program starts with Maxwell interpolation:

\begin{equation}
\frac{ds}{dt}=\frac{P}{\eta}+\frac{1}{G}\frac{dP}{dt}
\label{a1}
\end{equation}
\noindent where $s$ is the shear strain, $\eta$ is viscosity, $G$ is shear modulus and $P$ is the shear stress.

The relation (\ref{a1}) reflects Maxwell's proposal \cite{Maxwell01011867} that the shear response in a liquid is the sum of viscous and elastic responses given by the first and second right-hand side terms. Importantly (we will come to this point later), the dissipative term containing the viscosity is not introduced as a small perturbation: both elastic and viscous deformations are treated in (\ref{a1}) on equal footing.

Frenkel proposed \cite{doi:10.1021/j150454a025} to represent the Maxwell interpolation by introducing the operator $A$ as $A=1+\tau\frac{d}{dt}$ so that Eq. (\ref{a1}) can be written as $\frac{ds}{dt}=\frac{1}{\eta}AP$. Here, $\tau$ is Maxwell relaxation time 

\begin{equation}
\tau_{\rm M}=\frac{\eta}{G_{\infty}}
\label{taum}
\end{equation}
where $G_{\infty}$ is the elastic modulus at infinite frequency which, differently from the static one (at zero frequency), it is finite even in liquids.
At the microscopic level, Frenkel's theory approximately identifies $\tau_{\rm M}$ with the time between consecutive diffusive jumps in the liquid \cite{doi:10.1021/j150454a025}. This has become an accepted view since \cite{RevModPhys.78.953}. Frenkel's idea was to generalize $\eta$ to account for liquid's short-time elasticity as

\begin{equation}
\frac{1}{\eta}\rightarrow\frac{1}{\eta}\left(1+\tau\frac{d}{dt}\right) \label{sub}
\end{equation}

\noindent and subsequently use this $\eta$ in the Navier-Stokes equation

\begin{equation}
\nabla^2\,v\,=\,\frac{1}{\eta}\,\left(\rho \frac{dv}{dt}\,+\,\nabla p\right)
\end{equation}

\noindent where $v$ is velocity field, $\eta$ is viscosity, $\rho$ is density, $p$ is pressure and $d/dt=\partial_t+v \cdot \nabla$.

This gives

\begin{equation}
\eta\,\frac{\partial^2 v}{\partial x^2}\,=\,\left(1\,+\,\tau\,\frac{d}{dt}\right)\,\left(\rho\,\frac{dv}{dt}\,+\,\nabla p\right)\label{super}
\end{equation}

Frenkel did not discuss the implications of \eqref{super}. We can solve the previous equation assuming a slowly flowing fluid $d/dt=\partial_t$ \cite{2016RPPh...79a6502T} and neglecting the non-linear terms $\mathcal{O}(v^2)$. For the transverse component of the velocity, which is decoupled from the pressure fluctuations $\delta p$, we obtain the following simple equation:

\begin{equation}
\eta\,\frac{\partial^2 v}{\partial x^2}\,=\,\rho\,\tau\,\frac{\partial^2 v}{\partial t^2}\,+\,\rho\,\frac{\partial v}{\partial t}
\end{equation}
which, using $\eta=G \tau=\rho \mathcal{V}^2 \tau$, can be re-written as:
\begin{equation}
\mathcal{V}^2\,\frac{\partial^2 v}{\partial x^2}\,=\,\frac{\partial^2 v}{\partial t^2}\,+\,\frac{1}{\tau}\frac{\partial v}{\partial t}
\end{equation}
where $\mathcal{V}$ is the propagation speed of the wave.\
We note that this equation can also be obtained by starting with the solid-like elastic equation for the non-decaying wave and, using Maxwell interpolation (\ref{a1}), generalizing the shear modulus to include the viscous response \cite{PhysRevE.96.062134}. This shows that the hydrodynamic approach commonly applied to liquids \cite{hydro} is not a unique starting point and that the solid-like elastic approach is equally legitimate, implying an interesting symmetry of the liquid description.

Let's now Fourier decompose the velocity field as $v=v_0\,e^{i k x -i \omega t}$ in order to get:
\begin{equation}
\omega^2\,+\,\omega\,\frac{i}{\tau}\,-\,\mathcal{V}^2\,k^2\,=\,0\label{mm}
\end{equation}
The dispersion relation of the shear modes obtained from the previous equation reads:
\begin{equation}
\omega\,=\,-\,\frac{i}{2\,\tau}\,\pm\,\sqrt{\mathcal{V}^2\,k^2\,-\,\frac{1}{4\,\tau^2}}\label{kgapeq}
\end{equation}
and it gives what is known as the k-gap\footnote{Notice that at this stage momentum is a perfectly conserved quantity. We will analyze later how to modify this picture in presence of slow momentum relaxation.}. Notice that at this point we are considering non-relativistic systems. On the contrary when we will deal with our holographic models with an AdS UV asymptotic we will consider relativistic hydrodynamics and as a consequence we will have $\mathcal{V}=c$.

\begin{figure}[h]
\centering
\includegraphics[width=7.1cm]{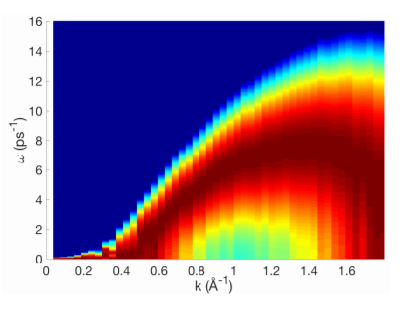}
\quad
\includegraphics[width=7.7cm]{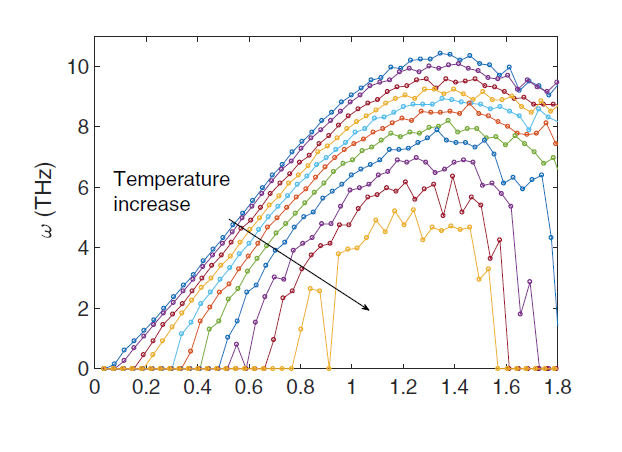}
\caption{Molecular dynamics simulations for supercritical Ar and CO2 between 200K and 500 K. The simulations show the presence of a k-gap in the transverse phonons dispersion relation. Figures taken from \cite{PhysRevLett.118.215502}.}
\label{th1}
\end{figure}

We observe that the appearance of the k-gap is a direct consequence of the unification of hydrodynamic and elasticity theories via Maxwell interpolation. In other words, the system in question (e.g. liquid) is capable of both hydrodynamic and solid-like elastic response. 

According to equation \eqref{kgapeq}, liquids can support propagating shear waves and therefore behave like solids even at low frequencies $\omega \approx 0$ as long as:
\begin{equation}
k\,>\,k_{gap}\,=\,\frac{1}{2\,\mathcal{V}\,\tau}
\label{kgap}
\end{equation}
We note that at large momenta $k \gg 1/\mathcal{V} \tau$ the dispersion relation approaches the linear one:
\begin{equation}
\omega\,=\,\mathcal{V}\,k
\end{equation}
which is typical of common solids, \textit{i.e.} massless transverse phonons. The same solid-like relationship follows when $\tau$ in (\ref{kgapeq}) becomes large, corresponding to the solid (recall that $\tau$ is the time between molecular jumps in the liquids). 

Recently \cite{PhysRevLett.118.215502}, detailed evidence for the $k$-gap was presented on the basis of molecular dynamics simulations (see Fig.\ref{th1}). It has been ascertained that $k_g$ increases with the inverse of liquid relaxation time in a wide range of temperature and pressure for different liquids and supercritical fluids, as (\ref{kgap}) predicts.
Moreover the $k$-gap represents a viable theoretical explanation for the detection of propagating shear waves at low frequency in liquids made in \cite{noirez2012identification}.\\

\begin{figure}
\centering
{\scalebox{0.45}{\includegraphics{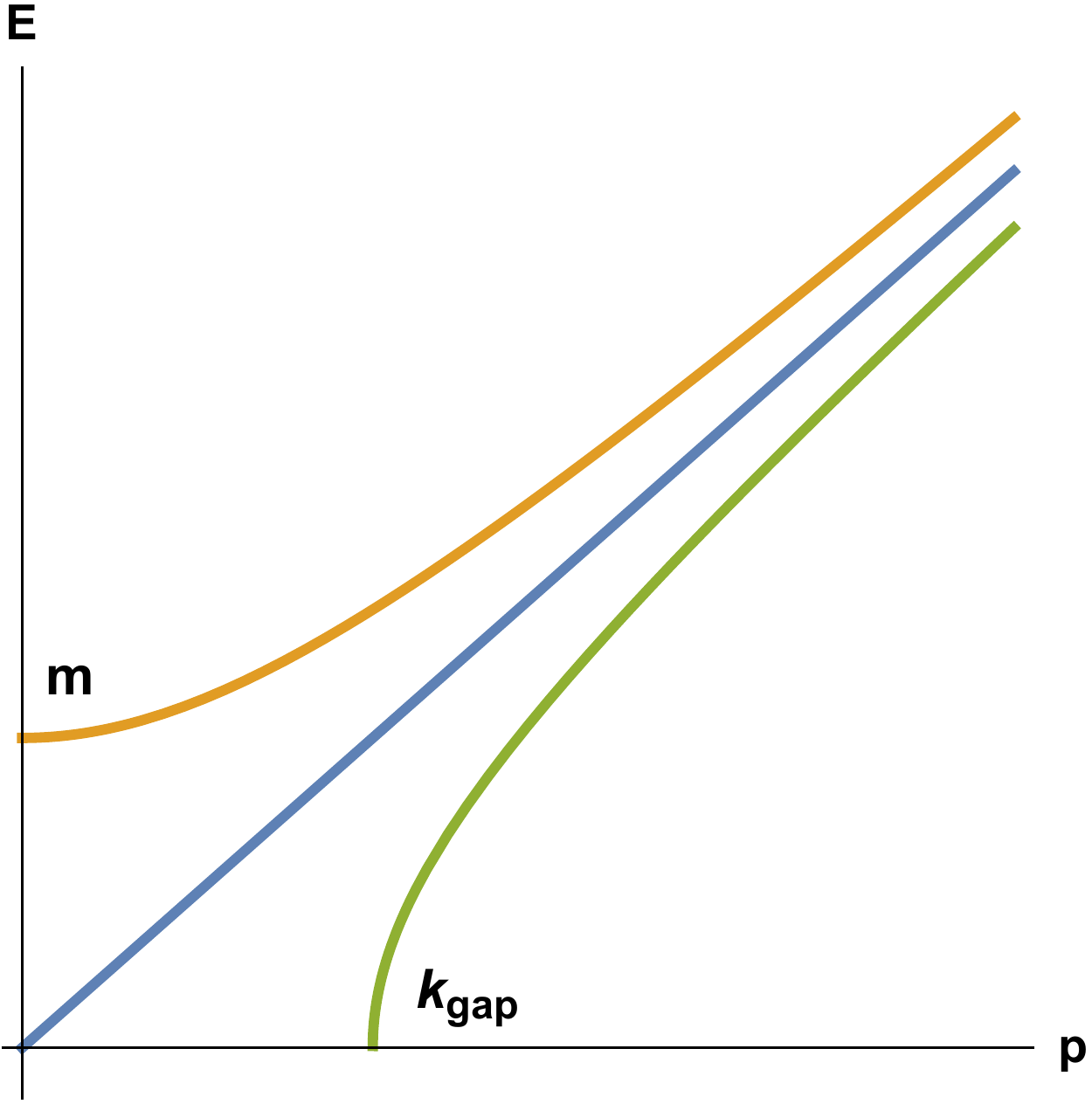}}}
\caption{Possible dispersion relations and dependencies of energy $E$ on momentum $p$. Top curve shows the dispersion relation for a massive particle. Middle curve shows gapless dispersion relation for a massless particle (photon) or a phonon in solids. Bottom curve shows the dispersion relation (\ref{ddd}) with the gap in $k$-space, illustrating the results of Ref. \cite{PhysRevLett.118.215502}.}
\label{3}
\end{figure}

The gap in $k$-space, or momentum space is interesting. Indeed, the two commonly discussed types of dispersion relations are either gapless as for photons and phonons, $E=p$ ($\mathcal{V}=c=1$), or have the energy (frequency) gap for massive particles, $E=\sqrt{p^2+m^2}$, where the gap is along the Y-axis. On the other hand, (\ref{ddd}) implies that the gap is in {\it momentum} space and along the X-axis, similar to the hypothesized tachyon particles with imaginary mass \cite{tachyons}. Figure \ref{3} illustrates this point.

Before proceeding, let us pause, and be clear about the Maxwell-Frenkel approach. There are indeed two steps: from one side Frenkel derived  the dispersion relation \eqref{kgapeq} without assuming any precise definition of the relaxation time $\tau$ but just using \eqref{sub}. On the other side Maxwell, using his well-known linear viscoelastic theory, proposed the definition of such relaxation timescale with the Maxwell relaxation time $\tau=\tau_M=\eta/G_{\infty}$. The two steps are completely independent and, as we will see in the following, the holographic models considered respect the dispersion relation form \eqref{kgapeq}, but their relaxation time is not  identifiable with the Maxwell relaxation time $\tau_M$.

Now let us consider a relativistic system described by hydrodynamic equations; in this case the asymptotic speed is simply the speed of light $\mathcal{V}=c$.
We can advance our analysis further and expand the expression \eqref{kgapeq} in the hydrodynamic limit $k \ll k_{gap}$:
\begin{equation}
\omega_1\,=\,-\,i\,c^2\,\tau\,k^2\,+\,\dots\,,\quad \omega_2\,=\,-\frac{i}{\tau}\,+\,i\,c^2\,\tau\,k^2\,+\,\dots \label{tt}
\end{equation}
From the latter expansion we can immediately realize that the relaxation time $\tau$ is related to the hydrodynamic diffusion constant\footnote{We assume the usual definition for a diffusive hydrodynamic mode $\omega=-i D k^2$.}
\begin{equation}
\boxed{D\,=\,c^2\,\tau} \label{fff}
\end{equation}
This is the main result of this section and the prediction we will compare with our numerical data from the holographic models.
This result can be derived from a quasihydrodynamic framework and it is valid as far as $\tau T \ll 1$.
Interestingly, if we assume the familiar expression for the diffusion constant in a relativistic fluid and the non relativistic speed of sound in solids:
\begin{equation}
D\,=\,\frac{\eta}{\chi_{PP}}\,,\quad c^2\,\rightarrow \mathcal{V}^2\,=\,\frac{G}{\chi_{PP}}
\end{equation}
we find 
\begin{equation}
\tau\,=\,\frac{\eta}{G}\label{pp}
\end{equation}
and, therefore, recover the result (\ref{taum}) suggested by Maxwell.

The presence of a k-gap in the shear spectrum of relativistic hydrodynamic and an implicit version of formula \eqref{fff} have already appeared in \cite{Pu:2009fj} within the discussion of the stability and causality of second order relativistic hydro.

We will generalize this framework for slow momentum relaxation in the next sections.\\

We note that a relation between the relaxation time and the diffusion constant is natural in condensed matter system\footnote{For example for Brownian motion that relation reads $D \sim a^2/\tau$ whith $a$ the lattice spacing.}. The main difference compared to gravity and holographic models is that in this case, due to the relativistic symmetries, the diffusion constant $D$ is directly, and not inversely, proportional to the viscosity of the system $\eta$ (see for more details, particularly related to Heavy Ions physics \cite{Kovtun:2012rj,Teaney:2009qa,CasalderreySolana:2011us}).
The proportionality between diffusion and viscosity corresponds to the high-temperature gas-like regime of particle dynamics in condensed matter where the oscillatory component of particle dynamics is lost. In this regime, relaxation time $\tau$ is related to the time between particle collisions, and the diffusion constant is proportional to viscosity as a result, with viscosity increasing with temperature \cite{doi:10.1021/j150454a025} as in holographic models. This behaviour is in contrast to low-temperature regime where the diffusion constant is inversely proportional to viscosity. Low- and high-temperature regimes of liquid dynamics are separated by the Frenkel line \cite{2016RPPh...79a6502T}.\\

In this work and the following sections we will provide evidence that the spectrum of liquids is strikingly similar to that following from simple bottom-up holographic models. The similarities are both general and detailed; we believe that the same general mechanism is at operation in these two very disparate areas and physical objects. 
\section{A simple holographic relative}\label{hol22}
In this section we consider a simple holographic model introduced in \cite{Grozdanov:2018ewh}, which displays interesting viscoelastic properties and the presence of a k-gap. Most of the computations presented here are already derived in the original paper. Nevertheless we will discuss all the phenomenological properties and we will show several interesting physical aspects of the model. Let us discuss the most relevant features of the setup.
\subsection{The model}
The model represents the holographic dual of a finite number of dynamical elastic lines defects embedded in a fluid state. The dynamical defects are described by two dynamical bulk two forms $B^I_{ab}$ whose field strengths $ H^I=dB^I$ are chosen to be:
\begin{equation}
H^1_{txr}\,=\,H^2_{tyr}\,=\,m
\end{equation}
The gravitational bulk action is defined as:
\begin{equation}
S\,=\,\frac{1}{2\,\kappa_4^2}\int\,d^4x\,\sqrt{g}\,\left(R\,+\,\frac{6}{L^2}\,-\,\frac{1}{12}\,H^I_{abc}H_I^{abc}\right)
\end{equation}
The temperature of the background is given by:
\begin{equation}
T\,=\,\frac{r_h}{4\,\pi}\,\left(3\,-\,\frac{m^2}{2\,r_h^2}\right)
\end{equation}
where $r_h$ is the position of the event horizon.
The parameter $m$ counts the number density of the line defects immersed in the fluid and accounts for the solid properties of the system. In particular the shear elastic modulus and the corresponding speed of transverse phonons are determined by $m$. For more details about the model we refer to \cite{Grozdanov:2018ewh}.\\
The setup at hand present two very important features:
\begin{enumerate}
\item No explicit breaking of translational invariance is present in the system. As a consequence the additional dissipative contribution caused by momentum relaxation is absent; the corresponding relaxation time is infinite $\tau_{rel}=\infty$. The lowest hydrodynamic mode, at small momenta $k/T \ll 1$, is therefore purely diffusive in agreement with what found in eq.\eqref{tt}. All in all the dispersion relation of the collective modes present in the $JJ$ correlator is \underline{exactly} identical to the one obtained from Frenkel reduction \eqref{mm}. In this case there is no need of using any generalization of the framework (but there will in our second model) but we will simply have $\tau=D/v^2$ where $v$ is the asymptotic speed of the transverse shear waves at large momenta.
\item The model contains a UV cutoff mass scale $\mathcal{M}$ which technically arises from the holographic renormalization procedure. This may correspond physically to the height of the interaction potential energy barrier $U$ present in liquids. See \cite{kostya} for more details.
\end{enumerate}
\subsection{The shear spectrum and the k-gap}
In this model the vibrational spectrum of the system, containing the phononic degrees of freedom, is encoded in the physics of the current associated with the bulk two-form. As a consequence if we want to look at the transverse shear modes we have this time to consider the following Green function:
\begin{figure}
\centering
\includegraphics[width=6.5cm]{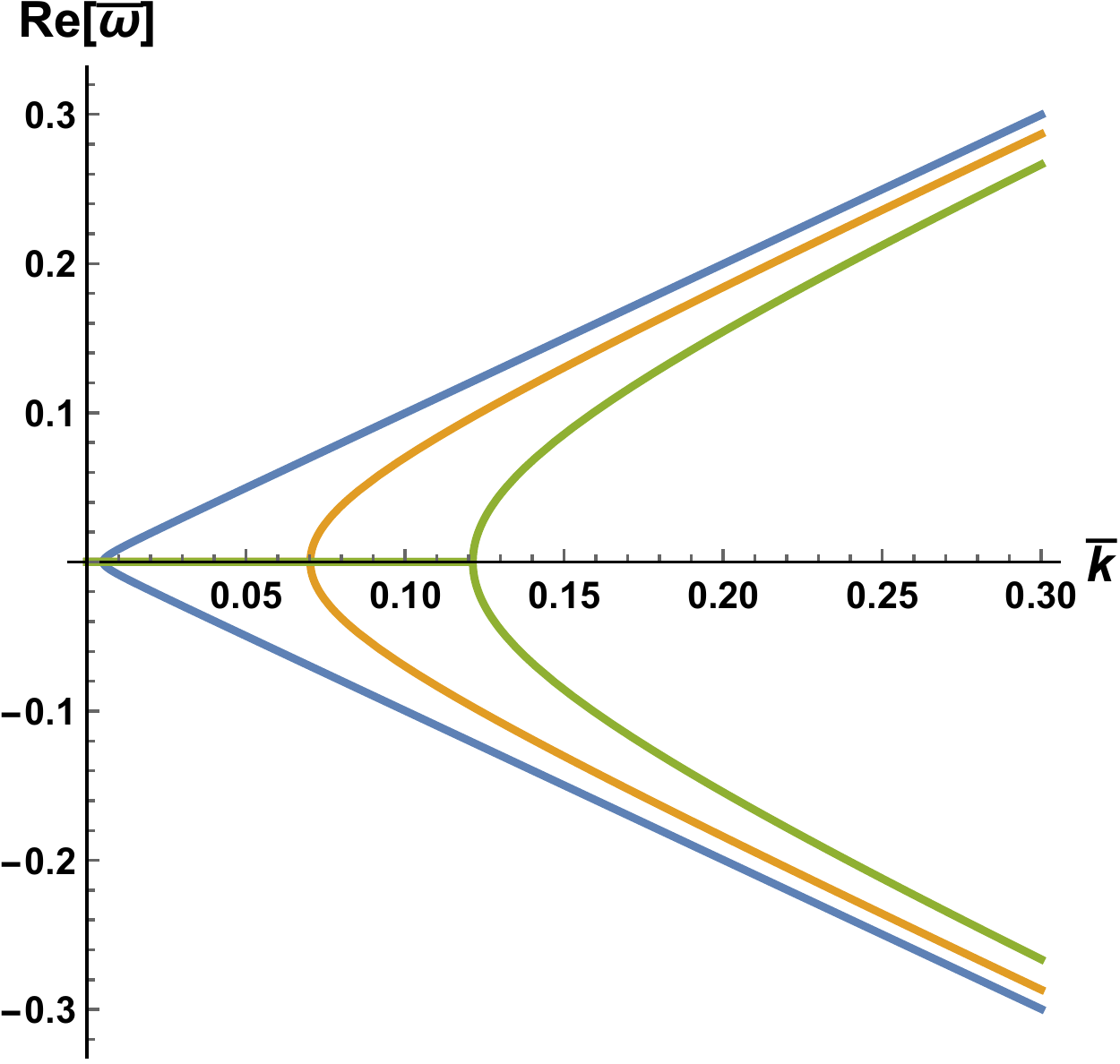}
\quad
\includegraphics[width=6.5cm]{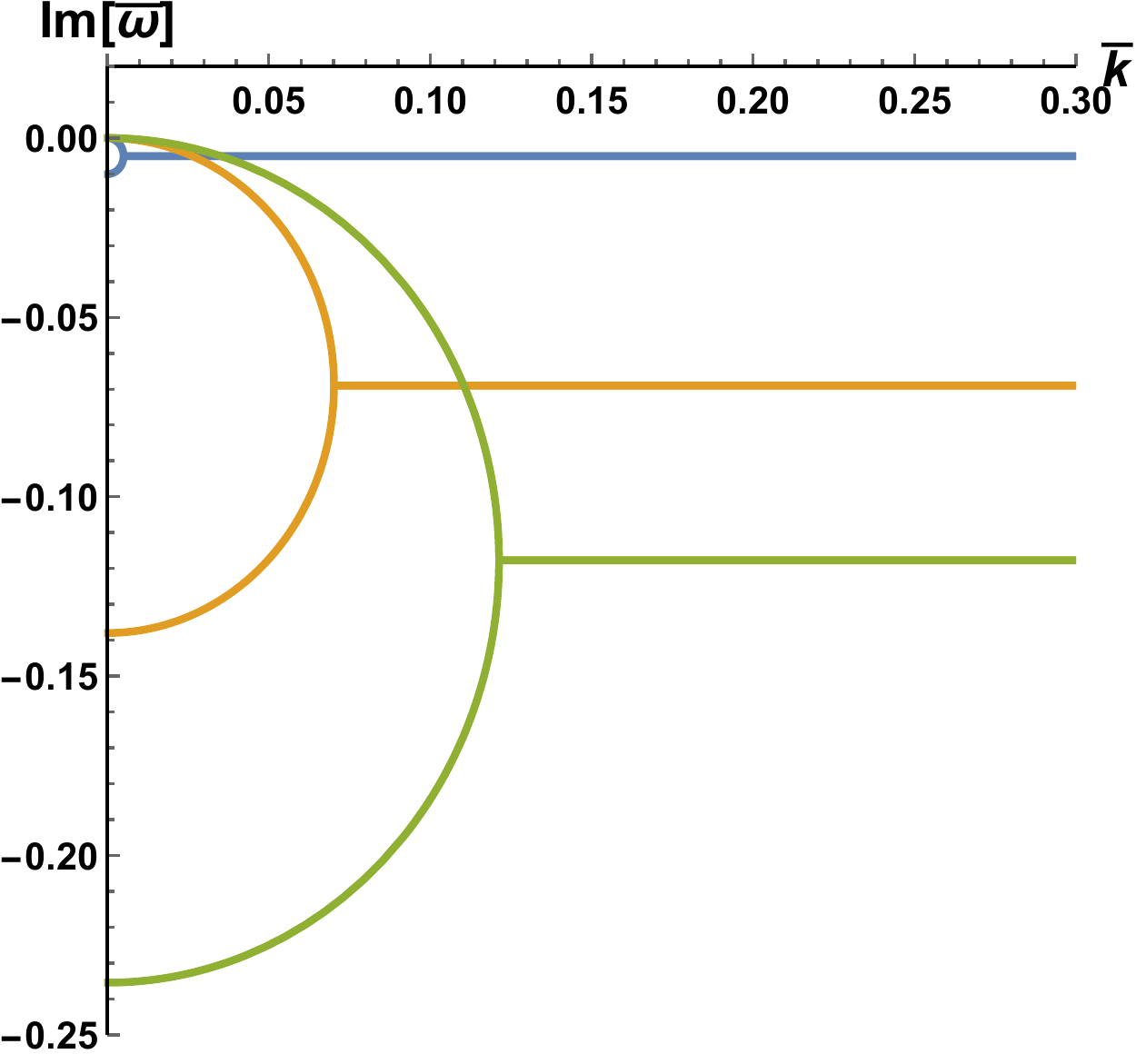}
\caption{The dispersion relation \eqref{disp} for $\mathcal{M}/r_h=5,8,100$ (from green to blue) in terms of the dimensionless frequency and momentum $\bar{\omega}=\omega/r_h$, $\bar{k}=k/r_h$. At large cutoff the $k$-gap closes and the dispersion relation becomes $\bar{\omega} = \bar{k}$.}
\label{fignick}
\end{figure}
Our main focus is the Green function of the current operators dual to the bulk two forms:
\begin{equation}
G^R_{JJ}(\omega,k)\,\equiv\,\langle J^I_{\mu\nu}J^I_{\sigma \rho}\rangle_R\,(\omega,k)
\end{equation}
which has not to be confused with the Green function of the electric current. In particular we are interested in the transverse part of such correlator that we will define $G^R_{TT,JJ}$ and in its poles, \textit{i.e.} transverse QNMs excitations.\\
In the following we focus on the case $m=0$ where no propagating transverse phonon is present at low frequency. In this limit the density of defects is zero and the system lies in a fluid (or better viscoelastic) phase. In the hydrodynamic limit $\omega/T \ll1,\,k/T \ll 1$ and in the limit of large enough cutoff $\mathcal{M}/T\gg 1$ the transverse QNMs are obtained solving the following equation\footnote{This assumes that $m=0$ and therefore the equations for the perturbations are not coupled.}
\begin{equation}
\left(1\,-\,i\,\frac{\omega}{\omega_g}\right)\omega\,+\,i\,\frac{\bar{M}\,-\,1}{r_h}\,k^2\,=\,0 \label{mmode}
\end{equation}
where $\bar{M}\equiv \mathcal{M}/r_h$. \\

\begin{figure}[h]
\centering
\includegraphics[width=9cm]{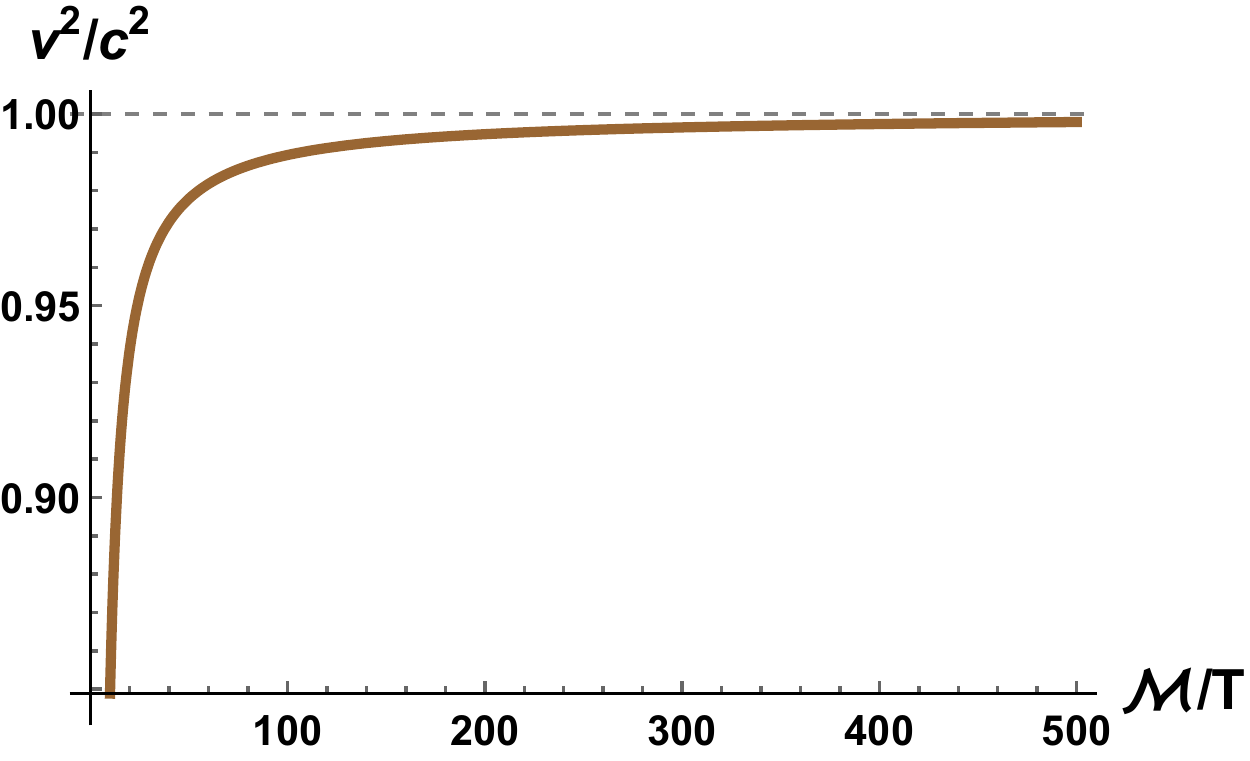}
\caption{The shear modes speed \eqref{fast} normalized to the relativistic speed $c=1$ in function of the cutoff $\mathcal{M}/T$.}
\label{figspeed}
\end{figure}

Strikingly, the above equation for the dispersion relation of the shear modes is exactly what found in  \cite{2016RPPh...79a6502T} which follows from Maxwell interpolation results \eqref{mm}.\\
Within this approximation we can derive the perturbative expression:
\begin{equation}
\omega_g\,=\,\frac{r_h}{\bar{M}-1+\frac{1}{2}(\ln 3-\frac{\pi}{3 \sqrt{3}})}
\end{equation}
Moreover, we can easily solve equation \eqref{mmode} and get the two dominating modes:
\begin{equation}
\omega_{\pm}\,=\,-\,\frac{\omega_g}{2}\,i\,\pm \sqrt{\frac{(\bar{M}-1)\,\omega_g}{r_h}\,k^2\,-\,\frac{\omega_g^2}{4}}\label{disp}
\end{equation}
The dispersion relation of such modes is of the type:
\begin{equation}
\omega\,=\,-\,\frac{i}{2\,\tau}\,\pm\,\sqrt{c^2\,k^2\,-\,\frac{1}{4\,\tau^2}}
\end{equation}
and exhibits a k-gap at:
\begin{equation}
k_{g}^2\,=\,\frac{\omega_g\,r_h}{4\,(\bar{M}-1)}\label{kgapnick}
\end{equation}
Notice how the lowest mode has $Im[\omega]=0$ at zero momentum $k=0$ as a consequence of the absence of momentum dissipation. An example of the dispersion relation is shown in fig.\ref{fignick}.
Notice that for large cutoff $\mathcal{M}/T, \gg 1$ the k-gap closes and the dispersion relation becomes exactly linear and relativistic $\omega=k$.
Let us first analyze the speed $c$ which appears at large momentum. Differently from the previous case, the speed now depends non trivially on the UV cutoff:
\begin{equation}
v^2\,=\,\frac{(\bar{M}-1)\omega_g}{r_h}=\frac{9 (4 \pi -3 \tilde{M})}{2 \pi  \left(18+\sqrt{3} \pi -9 \log (3)\right)-27 \tilde{M}}\label{fast}
\end{equation}
where we have used $\tilde{M}\equiv \mathcal{M}/T$. The results are presented in fig.\ref{figspeed}.
As expected, if we send the UV cutoff to infinity $\tilde{M}\rightarrow \infty$, we recover the relativistic speed value $v=c=1$. This confirms indeed the picture that such a feature is due to the absence of any UV cutoff.\\
Notice that the speed entering the k-gap dispersion relation \eqref{kgapeq} has not necessarily to be the speed of light. See for example in the context of kinetic theory also \cite{Romatschke:2009im}.\\
We now turn to studying the behaviour of the k-gap \eqref{kgapnick} and the relaxation time $\tau$ in this model.
\begin{figure}
\centering
\includegraphics[width=8cm]{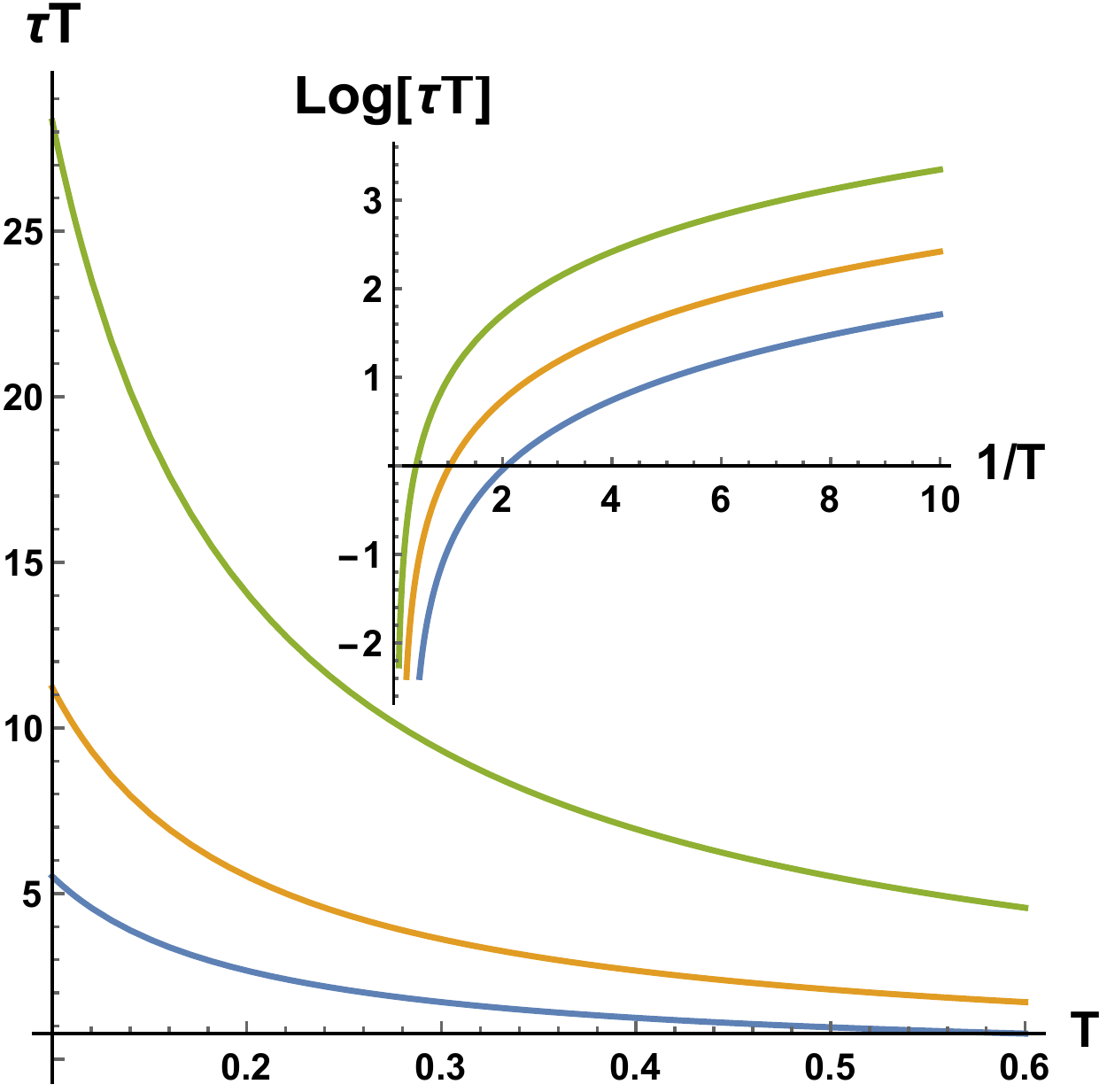}
\caption{The relaxation time $\tau$ in function of temperature at fixed $\mathcal{M}/T=10,20,50$ (from blue to green).}
\label{taunick}
\end{figure}
It can be immediately seen that:
\begin{equation}
\tau\,\equiv \frac{1}{\omega_g}\,=\,\frac{27 \,\tilde{M}\,-\,2\, \pi  \left(18+\sqrt{3} \pi -9 \log (3)\right)}{48 \,\pi ^2\, T}\label{tree}
\end{equation}
where $\tilde{M}\equiv M/T$. The first important feature to notice is that in the limit $\tilde{M}\rightarrow \infty$, the k-gap completely disappears. In another words, the existence of the k-gap in this model is fully dependent on the existence of a UV cutoff. To this extent this model appears very different from the previous one which shows a k-gap also in absence of a UV cutoff.  Fixing $\tilde{M}$, the relaxation time has a simple $1/T$ dependence which is shown in fig.\ref{taunick} for various values of the cutoff.

It is now interesting to ask whether the relaxation time can be explained in terms of the Maxwell relaxation time and/or if it is given in terms of the diffusion constant of the hydrodynamic mode.

Since the physics of the two forms sector is completely decoupled from the dynamics of the stress tensor at $m=0$ it appears already clear that the reply to the first question is simply no.
Let us discuss the physics of the transverse spectrum further. At low momenta the hydro pole in the $JJ$ correlator is purely diffusive and reads:
\begin{align}
&\omega\,=\,-\,i\,D_J\,k^2\,=\,-\,i\,\frac{3}{4\,\pi\,T}\,\left(\frac{3\,\mathcal{M}}{4\,\pi\,T}\,-\,1\right)\,k^2\,+\,\dots
\end{align}
It is important to underline that at $m=0$ the static shear modulus is zero and therefore the relaxation time $\tau$ can clearly not given in terms of it. This is another indication that the simplistic Maxwell relaxation time interpolation is not at work in this class of models. Clearly the shear viscosity and the elastic modulus cannot be the ingredient defining the relaxation time $\tau$, because the dynamics of the stress tensor where those two quantities ''live'' is completely decoupled at $m=0$ from the $J$ sector where $\tau$ appears.\\
It is still interesting to check if, as suggested in the previous sections, the relaxation time is given by:
\begin{equation}
\boxed{\tau\,=\,\frac{D_J}{v^2}}\label{test}
\end{equation}
Putting together the various expressions we obtain:
\begin{align}
&\frac{D_J}{v^2}\,=\,\frac{27 \,\tilde{M}-2 \,\pi\, \left(18+\sqrt{3}\, \pi -9 \,\log (3)\right)}{48 \,\pi ^2\, T}\label{due}
\end{align}
where again $\tilde{M}=M/T$.
This indeed proves that the expected theoretical picture of equation \eqref{test} is correct.\\
Finally, we compare the $k_{gap}$ \eqref{kgapnick} with the relaxation time $\tau=1/\omega_g$ in fig.\ref{last} for various cutoff values. At large relaxation times $1/\tau \ll 1$ the curve is quite linear. Decreasing the value of the relaxation time we observe a divergence from the linear behaviour which is more evident at small cutoff. The results are compatible with what observed in real liquids (see for example \cite{2016RPPh...79a6502T}).
\begin{figure}
\centering
\includegraphics[width=8cm]{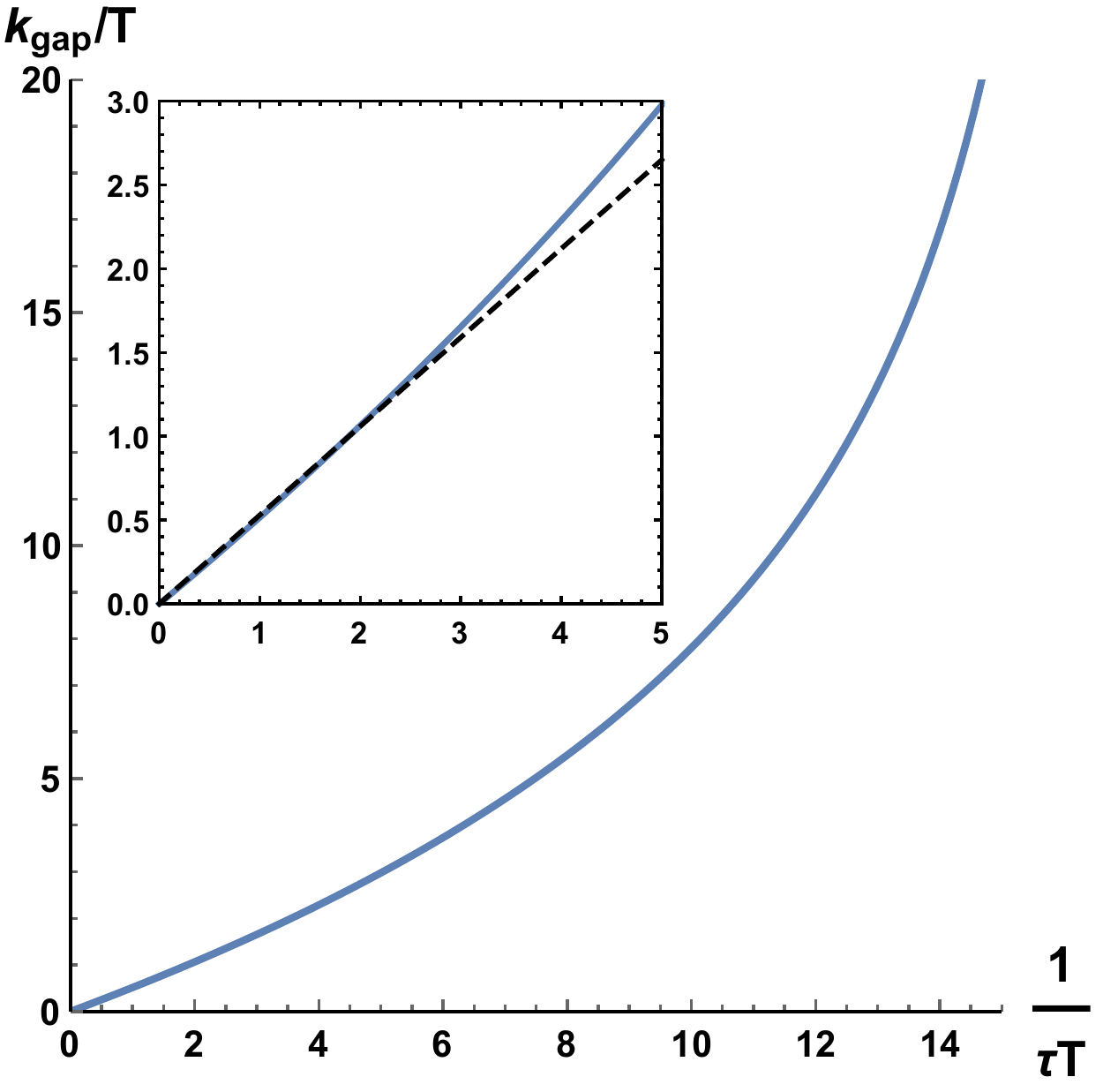}
\caption{The dimensionless $k_{gap}$ \eqref{kgapnick} in function of the dimensionless inverse relaxation time $1/T \tau$ for a large cutoff value $\tilde{M}=1000$. The inset shows the linear relation for small values of those quantities.}
\label{last}
\end{figure}
\section{A second holographic relative}\label{hol1}
In this section we present another holographic model which was introduced to realize momentum dissipation in the dual field theory. The model does not exhibit any shear elastic property at zero frequency and momentum, nor any propagating transverse phonons. In the hydrodynamic limit, it behaves like a fluid with a finite momentum relaxation time $\tau_{rel}$ where the dominant excitations are just damped diffusive modes. Despite numerous studies related to this model a deep analysis of the transverse modes away from the hydrodynamic limit was missing. In the following we will perform such a task and we will prove the existence of a k-gap in the dispersion relation for shear modes.
\subsection{The model}
We start considering a more generic class of holographic massive gravity models \cite{Baggioli:2014roa,Alberte:2015isw}:
\begin{equation}\label{S}
S\,=\, M_P^2\int d^4x \sqrt{-g}
\left[\frac{R}2+\frac{3}{\ell^2}- \, m^2 V(X)\right]
\end{equation}
with $X \equiv \frac12 \, g^{\mu\nu} \,\partial_\mu \phi^I \partial_\nu \phi^I$. These setups have been examined in various directions in the last years \cite{Baggioli:2015zoa,Baggioli:2015gsa,Baggioli:2016oqk,Baggioli:2016pia,Alberte:2017cch,Alberte:2017oqx} because they are particularly interesting from the condensed matter perspective. We study 4D AdS black brane geometries of the form:
\begin{equation}
\label{backg}
ds^2=\frac{\ell^2}{u^2} \left[\frac{du^2}{f(u)} -f(u)\,dt^2 + dx^2+dy^2\right] ~,
\end{equation}
where $u\in [0,u_h]$ is the radial holographic direction spanning from the boundary to the horizon, defined through $f(u_h)=0$, and $\ell$ is the AdS radius.\\

The $\phi^I$ fields are the St\"uckelberg scalars and they admit a radially constant profile $\phi^I=x^I$ with $I=x,y$.
This is an exact solution of the system due to the shift of symmetry. In the dual picture these fields represent scalar operators which break translational invariance because of their explicit dependence on the spatial coordinates.
Within these theories the helicity-2 metric perturbations acquire a mass term given by $m^2_g(u) = 2m^2 X\, V'$
with the background value for $X=u^2/\ell^2$. This is the reason why we refer to them as massive gravity theories.\\
The solution for the emblackening factor $f$ for generic potentials $V(X)$ is then given by
\begin{equation}\label{backf}
f(u)= u^3 \int_u^{u_h} dv\;\left[ \frac{3}{v^4} -\frac{m^2}{v^4}\,
V(v^2) \right] \, ,
\end{equation}
where $u_h$ stands for the location of the black brane horizon. The temperature and the entropy density are defined as
\begin{equation}
T=-\frac{f'(u_h)}{4\pi}=\frac{6  -  2 m^2 V\left(u_h^2 \right) }{8 \pi u_h}~,\quad s\,=\,\frac{2\pi}{u_h^2}\,.
\end{equation}\\
Finally the energy density of the dual field theory reads
\begin{equation}
\epsilon =\frac{1}{u_h^3}\left[1+\frac{m^2 \,  u_h^{2n}}{2n-3}\right]  \,,\qquad V(X)\,=\,X^n\,.
\end{equation}
and can be easily derived from \eqref{backf} or from the renormalized boundary stress tensor.
\begin{figure}
\centering
\includegraphics[width=8cm]{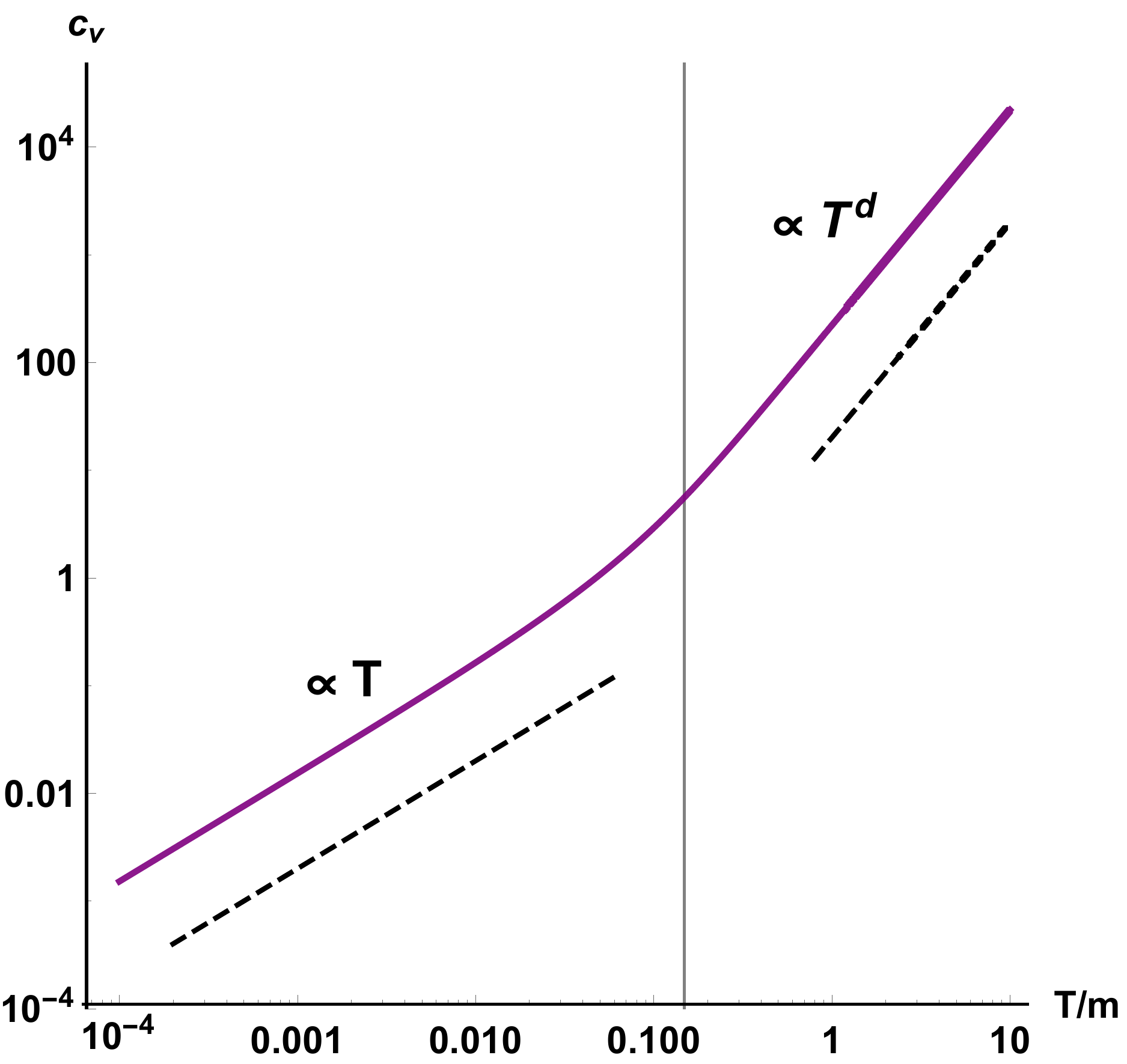}
\caption{The heat capacity in function of the dimensionless parameter $T/m$. The low temperature behaviour is $c_v \propto T$ while the high temperature limit is $c_v \propto T^d$ where in our case $d=2$. The crossover happens approximately at the value of $T/m\approx 0.148$ which coincides with the onset of the incoherent-coherent transition \cite{Davison:2014lua}.}
\label{figheat}
\end{figure}
The transverse or shear perturbations are then encoded in the fluctuations $a_x,\,h_{tx}\equiv u^2 \delta g_{tx},\,h_{xy}\equiv u^2\delta g_{xy},\,\delta \phi_x,\,\delta g_{xu}$. Assuming the radial gauge, \textit{i.e.} $\delta g_{xu}=0$, and using the ingoing Eddington-Finkelstein coordinates
\begin{equation}
ds^2=\frac{1}{u^2} \left[-f(u)dt^2-2\,dt\,du + dx^2+dy^2\right]
\end{equation}
the remaining equations read\footnote{Here we set the momentum $k$ along the $y$ direction. This freedom comes from the fact that the model enjoy SO(2) symmetry in the $(x,y)$ subspace.}
\begin{align}
&-2(1-u^2\,V''/V')h_{tx}+u\,h_{tx}'-i\,k\,u\, h_{xy}-\left( k^2\,u+2\,i\,\omega (1-u^2\,V''/V')\right)\,\delta \phi_x+u\,f\, \delta \phi_x''
\nonumber\\&+\left(-2(1-u^2\,V''/V')\,f+u\,(2 i \omega+f')\right)\delta \phi_x'\,=\,0
\,\,;\nonumber\\
&2\,i\,m^2\,u^{2}\omega V'\,\delta \phi_x+u^2\, k\,\omega\, h_{xy}+\left(6+k^2\,u^2-2 \, m^2 (V-u^2\,V' ) \, -6 f+2uf'\right) \, h_{tx}
\nonumber\\&+\left(2\,u\,f-i\,u^2\omega\right)h_{tx}'-u^2\,f\,h_{tx}''\,=\,0 \,\,;
\nonumber\\
&2i\,k\,u\,h_{tx}-iku^2h_{tx}'-2\,i \, k \,m^2 \,u^{2}V'\delta \phi_x+2 h_{xy}\left(3+i\,u \, \omega-3f+uf'- \,m^2(V-u^2 V') \right)
\nonumber\\&-\left(2i\,u^2 \, \omega-2uf+u^2\,f'\right)h_{xy}'-u^2\,f\, h_{xy}''\,=\,0\,\,;
\nonumber\\
&2\,h_{tx}'-u\,h_{tx}''-2m^2\,u\,V' \, \delta\phi_x'+ik\,u\,h_{xy}'\,=\,0\,\,.
\end{align}
and they can be solved numerically.
The asymptotics of the various bulk fields close to the UV boundary $u=0$ are:
\begin{align}
&\delta \phi_x\,=\,\phi_{x\,(l)}\,(1\,+\,\dots)+\,\phi_{x\,(s)}\,u^{5-2n}\,(1\,+\,\dots)\,,\quad\nonumber\\ &h_{tx}\,=h_{tx\,(l)}\,(1\,+\,\dots)\,+\,h_{tx\,(s)}\,u^{3}\,(1\,+\,\dots)\,,\nonumber\\ &h_{xy}\,=h_{xy\,(l)}\,(1\,+\,\dots)\,+\,h_{xy\,(s)}\,u^{3}\,(1\,+\,\dots)\,.
\end{align}
In these coordinates the ingoing boundary conditions at the horizon are automatically satisfied by regularity at the horizon.
It follows that the various retarded Green's functions can be defined as:
\begin{align}\label{greenF}
&\mathcal{G}^{\textrm{(R)}}_{T_{tx}T_{tx}}\,=\,\frac{2\,\Delta-d}{2}\,\frac{h_{tx\,(s)}}{h_{tx\,(l)}}\,=\,\frac{3}{2}\frac{h_{tx\,(s)}}{h_{tx\,(l)}}\,,\nonumber\\
&\mathcal{G}^{\textrm{(R)}}_{T_{xy}T_{xy}}\,=\,\frac{2\,\Delta-d}{2}\,\frac{h_{xy\,(s)}}{h_{xy\,(l)}}\,=\,\frac{3}{2}\frac{h_{xy\,(s)}}{h_{xy\,(l)}}\,.
\end{align}
where spacetime dependences are omitted for simplicity. From the poles of the Green functions, defined as the zero of the leading terms in the UV expansions we can read off the QNMs frequency at finite momentum.
\\[0.2cm]
In this work we shall consider the simplest model
\begin{equation}
\label{X}
V(X)=X.
\end{equation}
which is usually referred to\footnote{This naming is quite misleading. The scalar fields present in this model are the Stueckelberg fields for the massive gravity theory at hand \cite{Alberte:2015isw}. They enjoy shift symmetry, as the axions, but their nature is completely different and very well known within the massive gravity community (see for example \cite{Hinterbichler:2011tt}).} as ''linear axions'' model \cite{Andrade:2013gsa}. The literature about this model is very vast and we do not intend to review it here. We will comment through the text about the main and relevant features of the model and the corresponding references.\\
For completeness let us remind the reader that this holographic model is thought to be an effective description for homogeneous momentum relaxation, where the momentum relaxation time is controlled by the bulk graviton mass. From the Condensed Matter perspective it might represent an averaged description for a disordered system whose features are very similar to the properties of amorphous solids.\\
Before proceeding let us just discuss some thermodynamic properties of the model. In particular we can derive the heat capacity as (see \cite{Baggioli:2015gsa} for previous discussions):
\begin{equation}
c_v\,=\,T\,\frac{ds}{dT}\,=\,\frac{4 \pi  \left(3\,-\,m^2\, u_h^2\right)}{u_h^2 \left(m^2
   \,u_h^2+3\right)}
\end{equation}
The result is shown in fig.\ref{figheat}. The heat capacity displays a crossover behaviour between a low temperature and a high temperature scaling. In particular for low T we have $c_v \propto T$ while at high T we have $c_v \propto T^d$ where in our case $d=2$. The crossover point happens around $T/m \approx 0.148$ and coincides exactly with the incoherent-coherent transition. From the holographic point of view these scalings are a direct manifestation of the nature of the UV and IR fixed points. More in details the $T^d$ scaling follows immediately from dimensional analysis around the AdS UV geometry, while the linear in $T$ contribution at low temperature is a typical feature of the AdS$_2 \times R^2$ near horizon geometry (see \cite{Blake:2016jnn})\footnote{Let us try, as an aside comment, to give a more physical interpretation of this crossover. At large $T/m$ the vibrational modes are just relativistic plane waves $\omega=k$ for which the corresponding density of states is $g(\omega)\sim \omega^2$, \textit{i.e.} the well known Debye result. As a consequence their contribution to the specific heat, which can be obtained with a specific integral over the density of state, is just the usual Debye term $T^d$. That said, at the incoherent-coherent transition (which is known in condensed matter as the Ioffe-Regel crossover) the modes stop to propagate and the physics becomes totally diffusive \cite{allen1999diffusons} like in a strongly disordered system such as a glass or an amorphous solid \cite{RevModPhys.78.953,ped} where indeed the specific heat has a linear $T$ dependence at low temperature. In other words the scalings of the specific heat can be understood just in terms of the relevant vibrational degrees of freedom of the system: at high $T/m$ sound modes while at low $T/m$ diffusive modes.}. 
\subsection{The shear channel}
\begin{figure}
\center
\includegraphics[width=8cm]{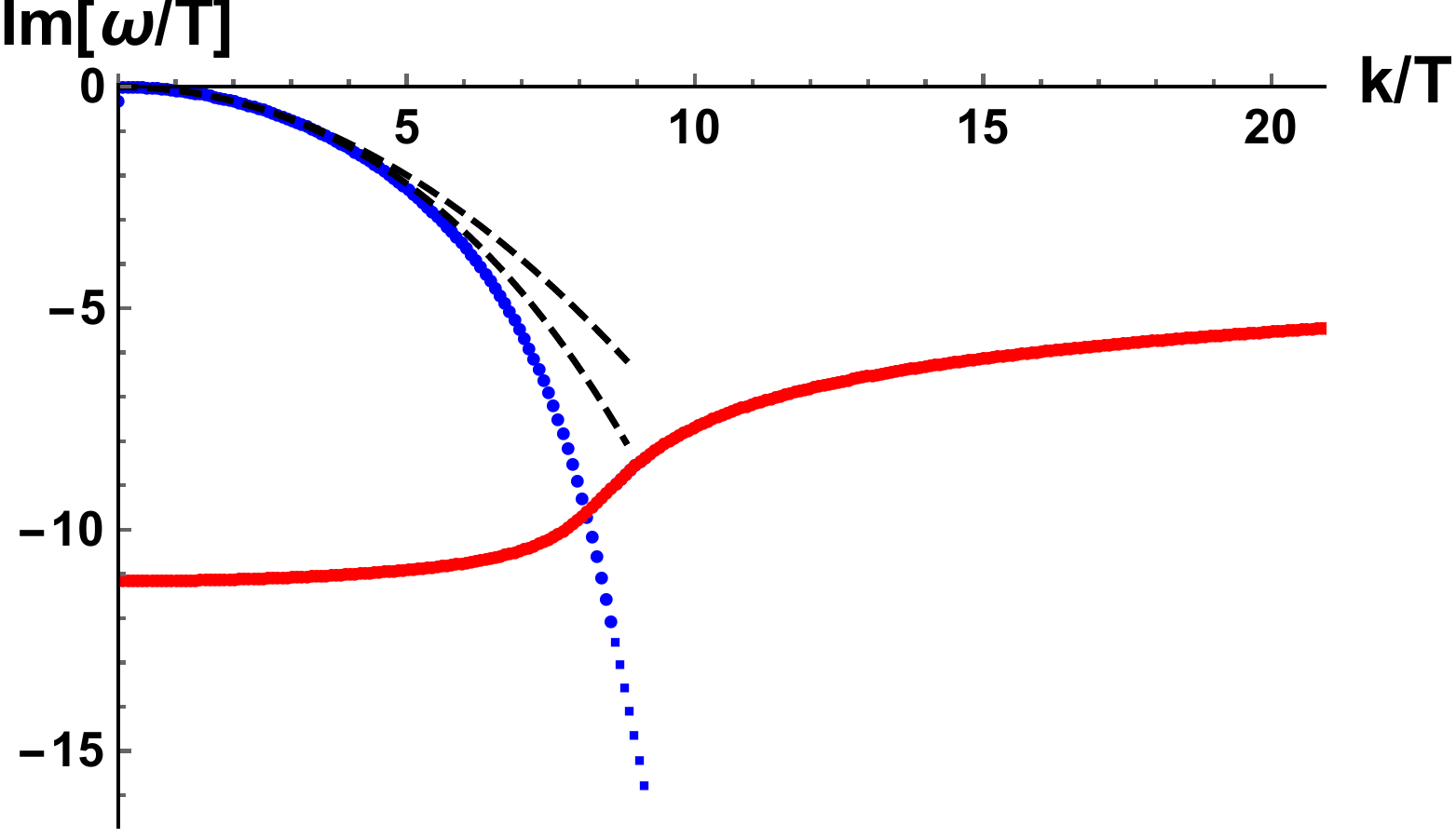}
\quad
\includegraphics[width=6.9cm]{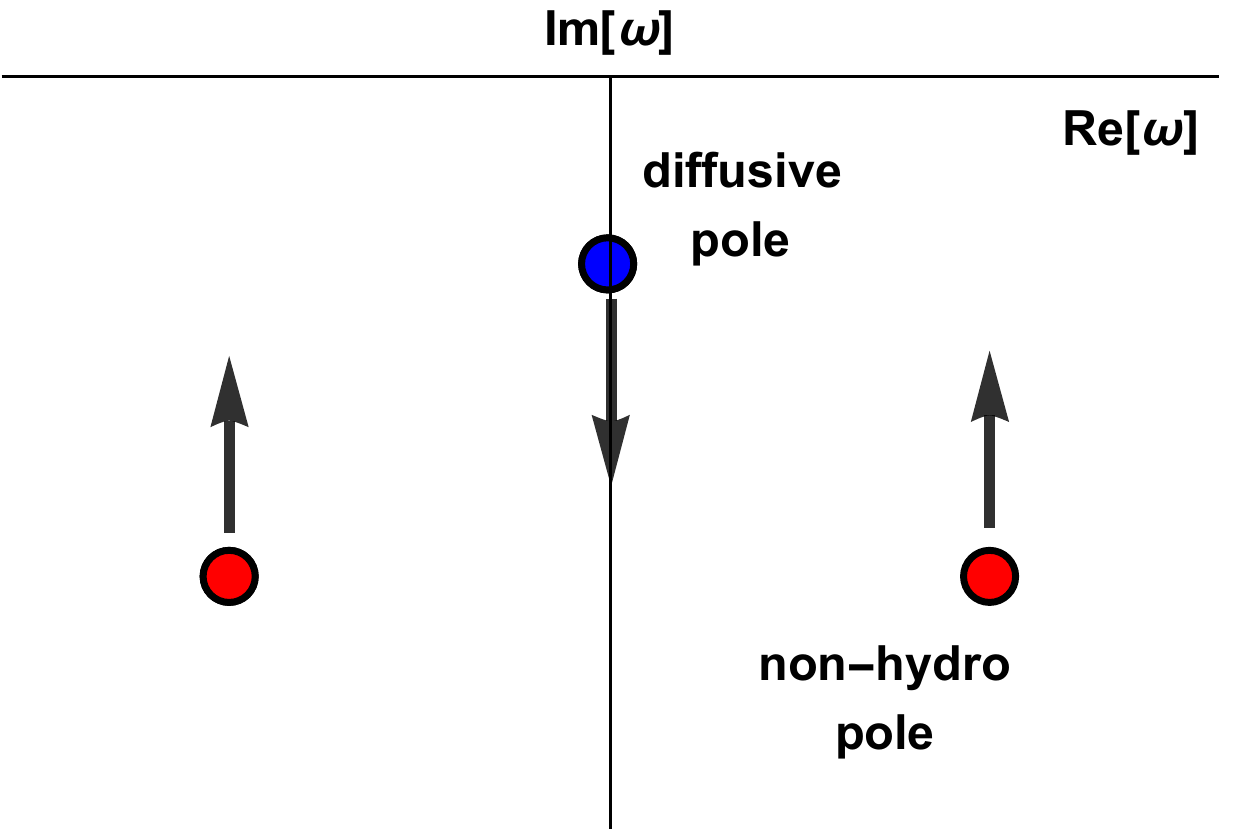}
\caption{The QNMs spectrum in the shear channel for $m^2=0$. \textbf{Left: }The dashed black lines are the analytic expression for the diffusion mode at $\mathcal{O}(k^2),\mathcal{O}(k^4)$ from \cite{Grozdanov:2015kqa}. The blue mode is the hydrodynamic diffusive one and the red one is the first non-hydro pole. Despite the crossing of the modes in the Imaginary axes there is no real modes collision in the complex plane and no $k$-gap is formed. \textbf{Right: }A cartoon of the poles crossover increasing $k/T$. This situation does not produce a $k-$gap but just a simple diffusion to sound crossover.}
\label{fig2}
\end{figure}
In this section we focus on the dynamical modes in the shear sector; we leave the longitudinal channel aside. For some complete discussion about Hydrodynamics, QNMs and BH fluctuations see \cite{Kovtun:2005ev,Nunez:2003eq,Policastro:2002se,Son:2002sd}.\\
In absence of any momentum relaxation time, \textit{i.e.} $m^2=0$, we are left with a pure Schwarzchild solution which corresponds to a relativistic hydrodynamic system \cite{Kovtun:2012rj}. In the hydro limit $\omega/T\ll1, k/T\ll 1$ the shear mode is purely diffusive:
\begin{equation}
\omega\,=\,-i\,D\,k^2\,-\,i\,\#\,k^4\,+\,\dots \label{diff1}
\end{equation}
where $\#$ is a specific combination of third order hydro coefficients \cite{Grozdanov:2015kqa} and $D$ is the momentum diffusion constant for a relativistic fluid:
\begin{equation}
D\,=\,\frac{\eta}{s\,T}\,=\,\frac{1}{4\,\pi\,T}
\end{equation}
Our interest is to study the spectrum of excitations of the system with $m^2 \neq 0$ and beyond the hydrodynamic limit.
We proceed with analyzing in more details the feature of the QNMs spectrum in the shear channel.\\
\begin{figure}
\center
\includegraphics[width=7.5cm]{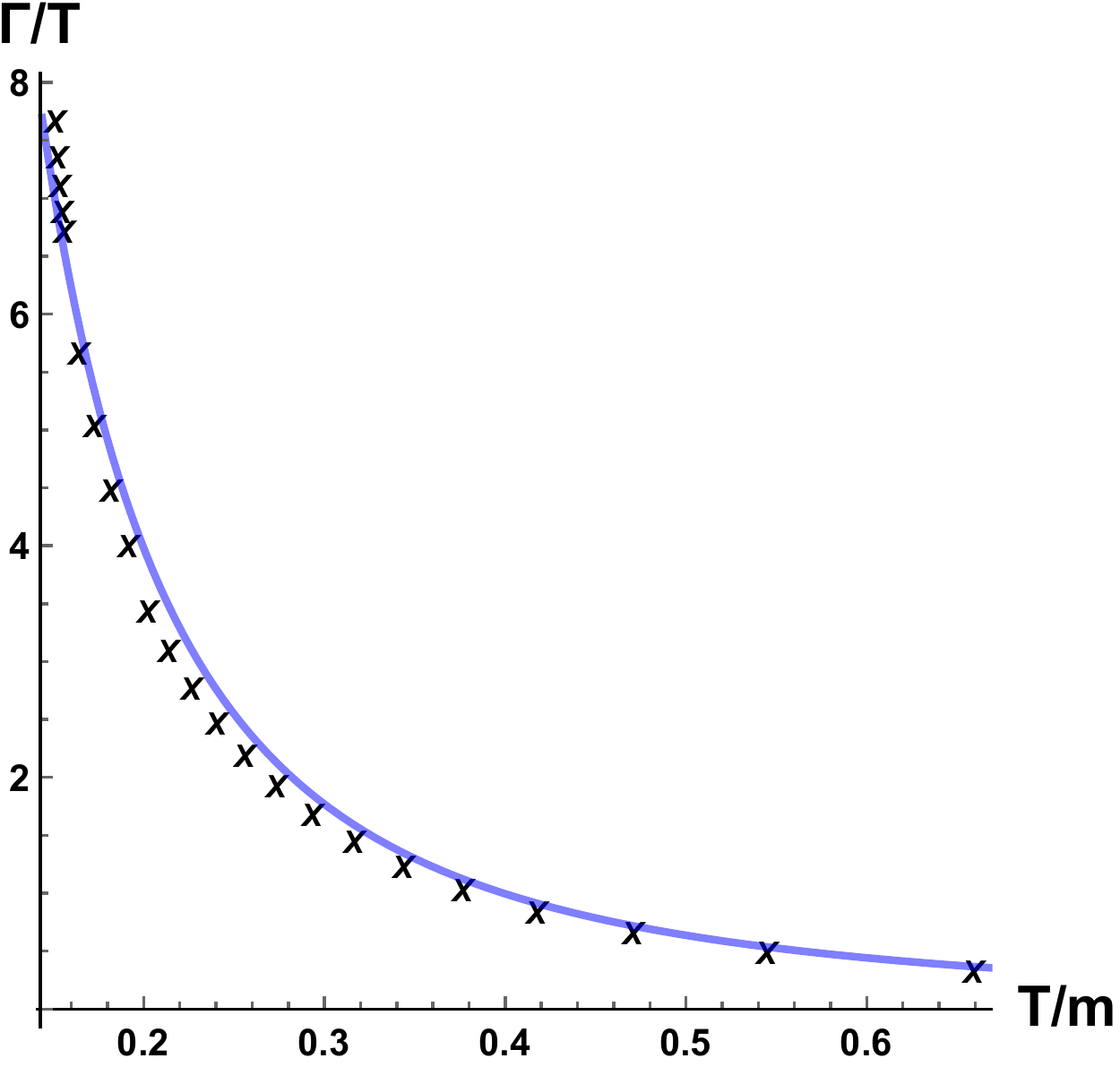}
\quad
\includegraphics[width=7.5cm]{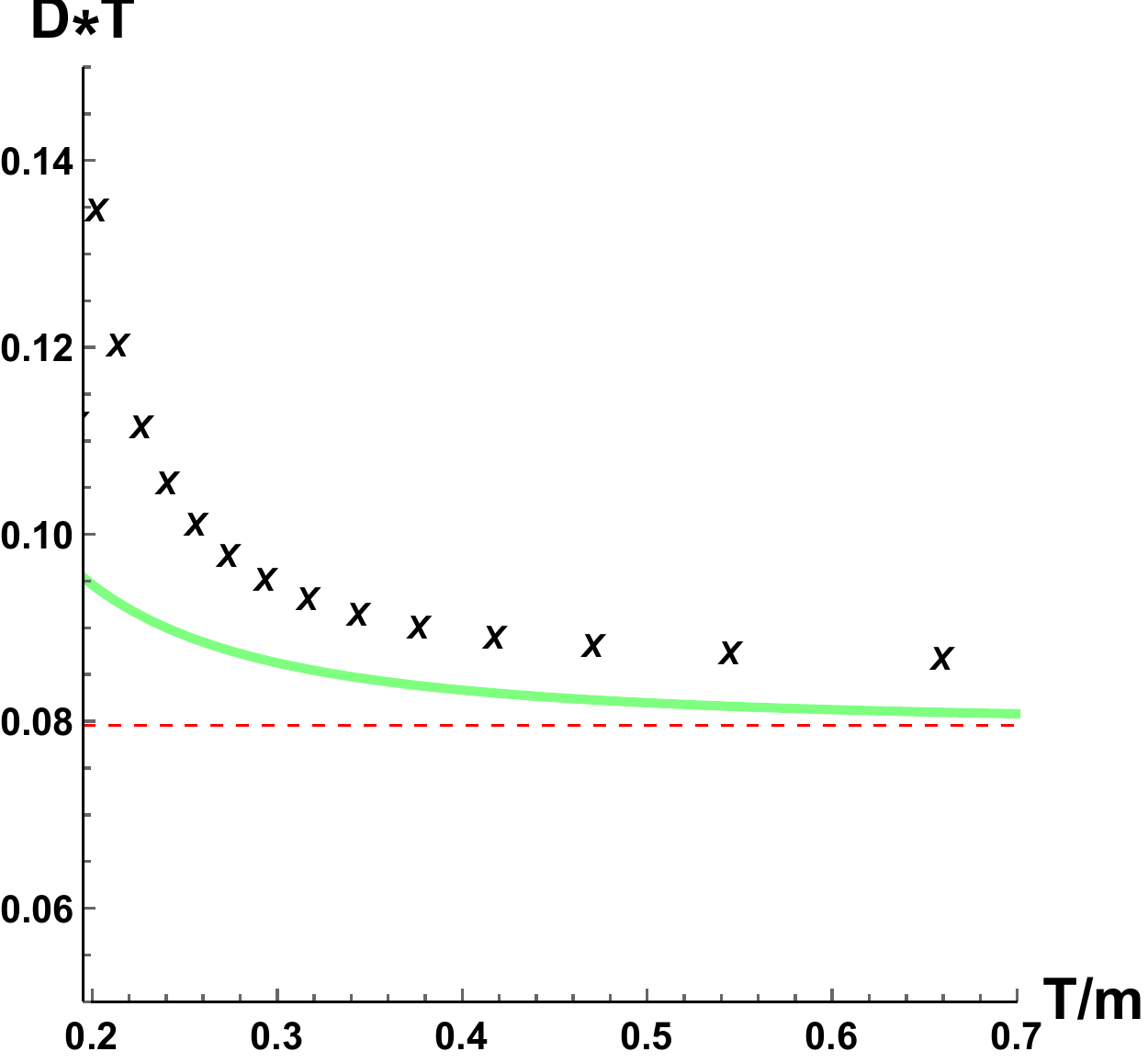}
\caption{\textbf{Left: }The relaxation time $\Gamma=\tau_{rel}^{-1}$ in function of temperature. The dots are the numerical values extracted from the QNMs while the purple curve is the approximated formula \eqref{approxtau} valid for $T/m \gg 1$. A similar plot can be found in \cite{Davison:2013jba,Amoretti:2014zha}. \textbf{Right: }The momentum diffusion constant $D$ extracted from the shear QNMs. The green curve is the approximated formula given in \eqref{anal}. The dashed line is the relativistic expression $DT=1/4\pi$ which turns out to be correct only at large temperature far from the regime drawn in this plot.}
\label{fig1}
\end{figure}
The zero mass case $m^2=0$ has been studied in details in various papers. For completeness we just repeat some salient features which will be useful for the future. The results are summarized in fig.\ref{fig2}. The only hydrodynamic mode in the shear channel is the diffusion mode \eqref{diff1} with exactly zero real part. Additional non-hydro modes are present in the spectrum. At a certain critical momentum $k^\star$ one of the initially overdamped mode crosses the diffusive mode and becomes the dominant one at large momentum $k \gg k^\star$. Moreover such a mode at large momentum asymptotes a relativistic dispersion relation $\omega=k$, forced by the UV relativistic fixed point, and in particular it becomes a propagating ''sound'' mode with $Re[\omega]\gg Im[\omega]$. The large momentum limit can be analyzed using the WKB methods \cite{Festuccia:2008zx,Fuini:2016qsc}. Nevertheless, for $m^2=0$ no k-gap appears in the spectrum and the phenomenology is far from the one discussed in the introduction. Indeed, despite the apparent crossing in the Imaginary axes there is no modes collision on the full complex plane.\\

We emphasize that the $k$-gap phenomenon is not just a simple diffusion to sound crossover, which indeed appears ubiquitously in almost every holographic system. Visualizing the $k$-gap in the complex frequency plane, it is just the manifestation of the collision (and not crossover) of two purely immaginary poles which on the contrary is not very common to happen in the transverse sector of the fluctuations.\\

Indeed, the situation changes drastically for non-zero mass. In the limit of small $m/T$, a new dimensionful scale, \textit{i.e.} the momentum relaxation time $\tau_{rel}^{-1} \propto m^2$, appears in the system and it competes with the original diffusion constant $D \propto \eta$. The diffusion mode becomes damped and acquires a finite imaginary part at zero momentum which sets the rate of dissipation for the momentum operator. Moreover, at zero momentum and a precise value of $m/T$ a coherent-incoherent transition appears and the hydrodynamic description in terms of a small relaxation time stops to be valid (see \cite{Davison:2014lua} for more details).\\
Let us start with analyzing the system within the hydrodynamic limit. The dispersion relation of the shear mode is now modified as:
\begin{equation}
\omega\,=\,-\,i\,\Gamma\,-\,i\,D\,k^2\,+\,\dots\label{pp}
\end{equation}
where $\Gamma=\tau_{rel}^{-1}$ controls the momentum relaxation timescale and $D$ is still the diffusion constant now $D \neq \eta/(sT)$. In order for the hydrodynamic expansion to be sensible we need the relaxation time to be not too short compared with the natural thermal timescale:
\begin{equation}
\tau_{rel}\,T\,\gg\,1
\end{equation}
Notice how the modified dispersion relation \eqref{pp} is not exactly the same of the k-gap equation \eqref{kgapeq} because of the presence of a finite relaxation time $\Gamma=\tau_{rel}^{-1}$ which is not included in the original Navier-Stokes equation in section \ref{prologue}.\\
\begin{figure}
\center
\includegraphics[width=7.5cm]{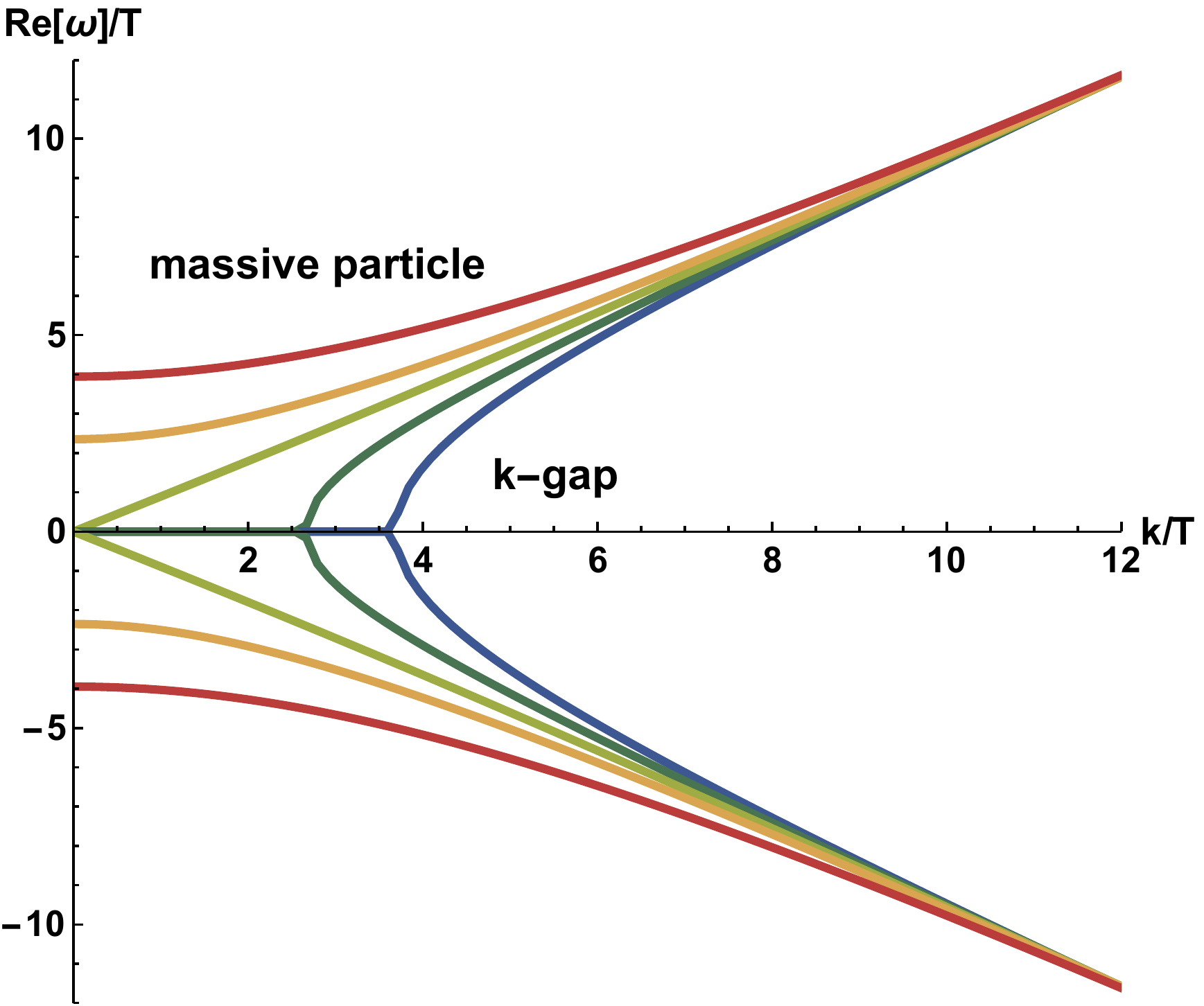}
\quad
\includegraphics[width=7.5cm]{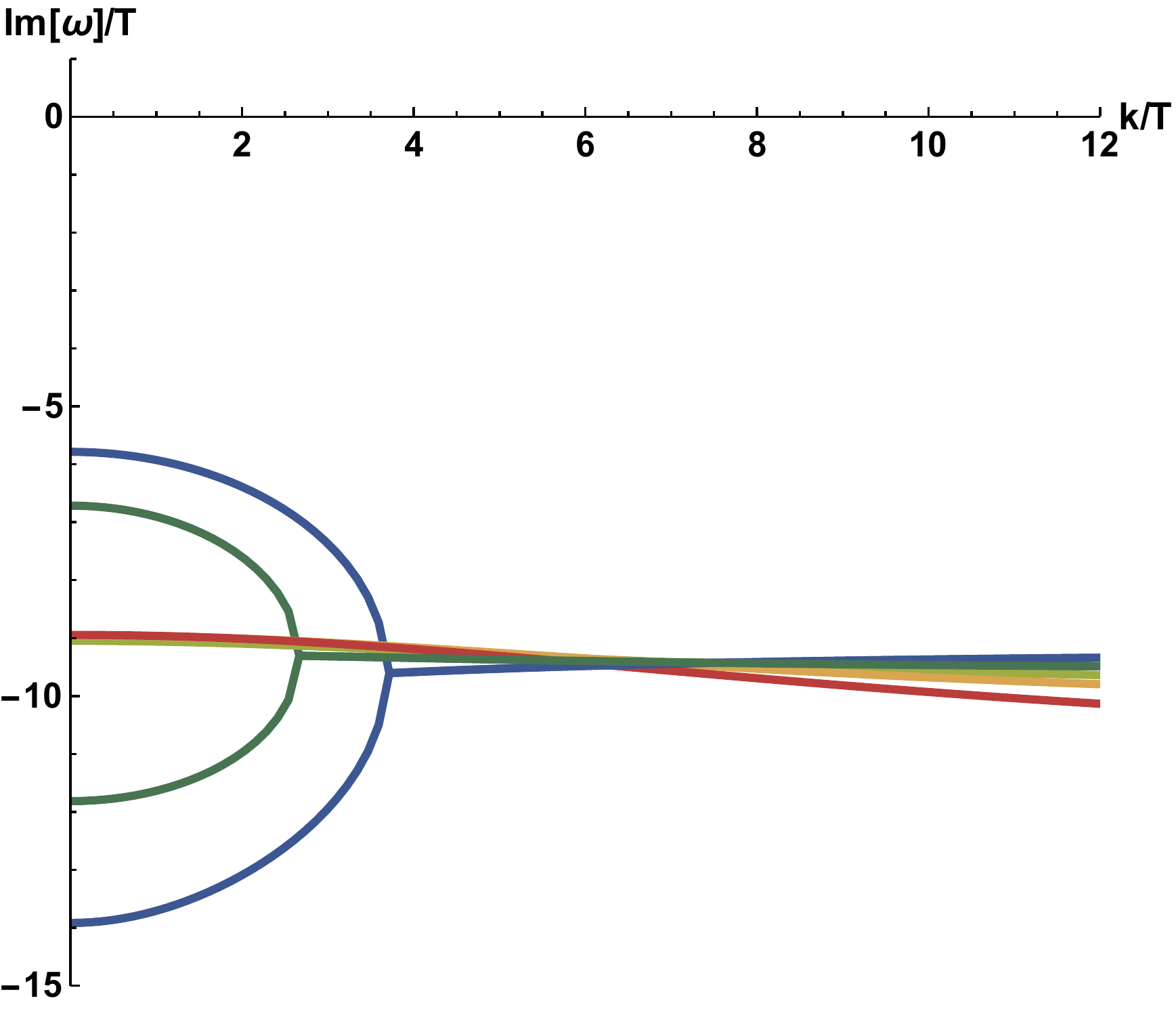}
\caption{The shear channel QNMs spectrum for various $T/m$ (from red to blue $T/m$ is increasing). The figure is adapted from \cite{Alberte:2017cch}.}
\label{fig3}
\end{figure}
For small $m/T$ we can write down the relaxation time as \cite{Davison:2013jba} :
\begin{equation}
\Gamma\,=\,\tau_{rel}^{-1}\,=\,\frac{m^2}{2\,\pi\,T}\,+\,\dots\label{approxtau}
\end{equation}
which implies that the hydro approximation is valid as far as $m/T \ll 1$. In the same limit we can have an approximate formula for the diffusion constant $D$ \cite{Ciobanu:2017fef} and for the viscosity of the system \cite{Hartnoll:2016tri,Alberte:2016xja,Burikham:2016roo}.
In particular the results of \cite{Ciobanu:2017fef} show that
\begin{equation}
D\,=\,\frac{1}{4\,\pi\,T}\,\left[1\,+\,\frac{1}{24}\,\left(9\,+\,\sqrt{3}\,\pi\,-\,9\,\log 3\right)\frac{m^2}{8\,\pi^2\,T^2}\right]\,+\,\dots \label{anal}
\end{equation}
in the limit of large temperatures, \textit{i.e.} $T/m \gg 1$, where the ellipsis stand for higher order terms in the dimensionless parameter $m/T$.
We plot the diffusion constant and the relaxation time for our system in fig.\ref{fig1}.
We notice that the approximate formulae are not a good approximation for $m/T \gg 1$, a range in which we could just rely on the numerics. We also notice that the diffusion constant of the system is no longer given by the usual relation:
\begin{equation}
D\,\neq\,\frac{\eta}{s\,T}
\end{equation}
and we believe that this is related to the fact that $\chi_{PP}=\epsilon+p\neq sT$\footnote{The relation between the diffusion constant of the shear hydrodynamic mode and the shear viscosity $D=\eta/\chi_{PP}$ is very generic and it does not rely on the assumption of relativistic invariance. In the case of a non-relativistic system $\chi_{PP}$ would just be the mass density of the system.} and in particular:
\begin{equation}
\chi_{PP}\,=\,\frac{3}{2}\,\epsilon\,=\,s\,T\,+\,2\,\mathcal{K}
\end{equation}
where $\mathcal{K}$ is the elastic bulk modulus. In different words the failure of the usual relation $D=\eta/sT$ is a consequence of the fact that the mechanical pressure and the thermodynamic one (extracted from the free energy) do not coincide for $m^2\neq 0$. This phenomenon has already been discussed and applied in \cite{Alberte:2017oqx,Alberte:2017cch}. A more detailed fluid gravity study is necessary to solve this puzzle \cite{progress}.\\
Moreover, decreasing the temperature $T$ the diffusion constant grows and the momentum relaxation time becomes smaller. Notice that in non-relativistic liquids the diffusion constant usually increases with temperature and is inversely proportional to the viscosity $\eta$. The difference hereby is caused by the relativistic symmetries of our system. Despite this difference, we will see that the temperature dependence of the $k$-gap and the dependence of the gap on relaxation time are the same in the two pictures: holographic gravity models and liquids.\\
At finite momentum $k\neq 0$ the situation is more complicated and depends crucially on $T/m$. Three different situations can arise and are summarized in fig.\ref{fig3} (notice, once again, the surprising similarities with fig.\ref{3} and the analysis of \cite{PhysRevE.96.062134}). For small $T/m$ the lowest hydrodynamic mode is gapped and shows a dispersion relation typical of a massive particle with mass $\mathfrak{m}$, $\omega^2\,=\,k^2\,+\,\mathfrak{m}^2$; moreover the mode is overdamped with an approximately constant imaginary part. Such modes share several features with the instantaneous normal modes observed and discussed in liquids \cite{PhysRevLett.83.108}. Nevertheless, it is not clear to us if this regime is sensible and this kind of dispersion relation is observed in any realistic situation. Notice also that within such a regime the naive energy density $\epsilon$ is negative. Increasing $T/m$ there is a particular value at which the dispersion relation becomes exactly linear $\omega=k$. Such a point signals the onset of incoherent-coherent transition at zero momentum which has been studied in the past literature in various contexts (see \cite{Davison:2014lua,Kim:2014bza}). Finally, increasing the temperature further, the physics is dominated by the hydro diffusive mode at low momenta $k<k_{gap}$. At a certain critical momentum $k\equiv k_{gap}$ the diffusive mode collides with the first non-hydro mode and a gap in momentum space opens. These features have been already studied from the QFT perspective in \cite{PhysRevE.96.062134} and will be of interest in this work.\\
Let us notice that in the limit $m/T \rightarrow 0$ the lowest of the two modes in fig.\ref{fig3} becomes more and more damped and it disappears from the hydrodynamic window and from the relevant late time dynamics. On the contrary the other pole goes forming the usual shear diffusion mode moving towards the origin of the complex plane.\\
\begin{figure}
\center
\includegraphics[width=7.5cm]{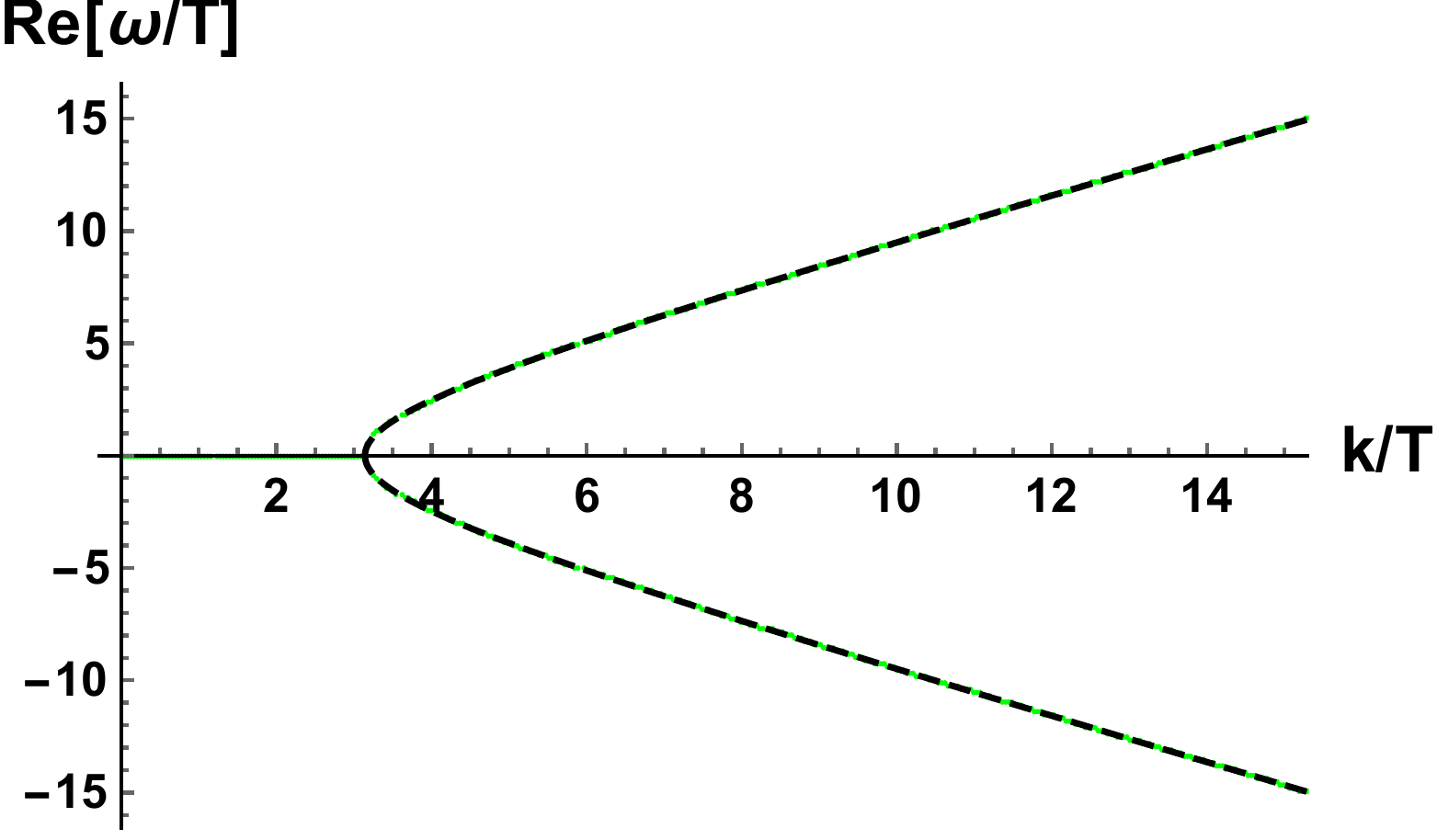}
\quad
\includegraphics[width=7.5cm]{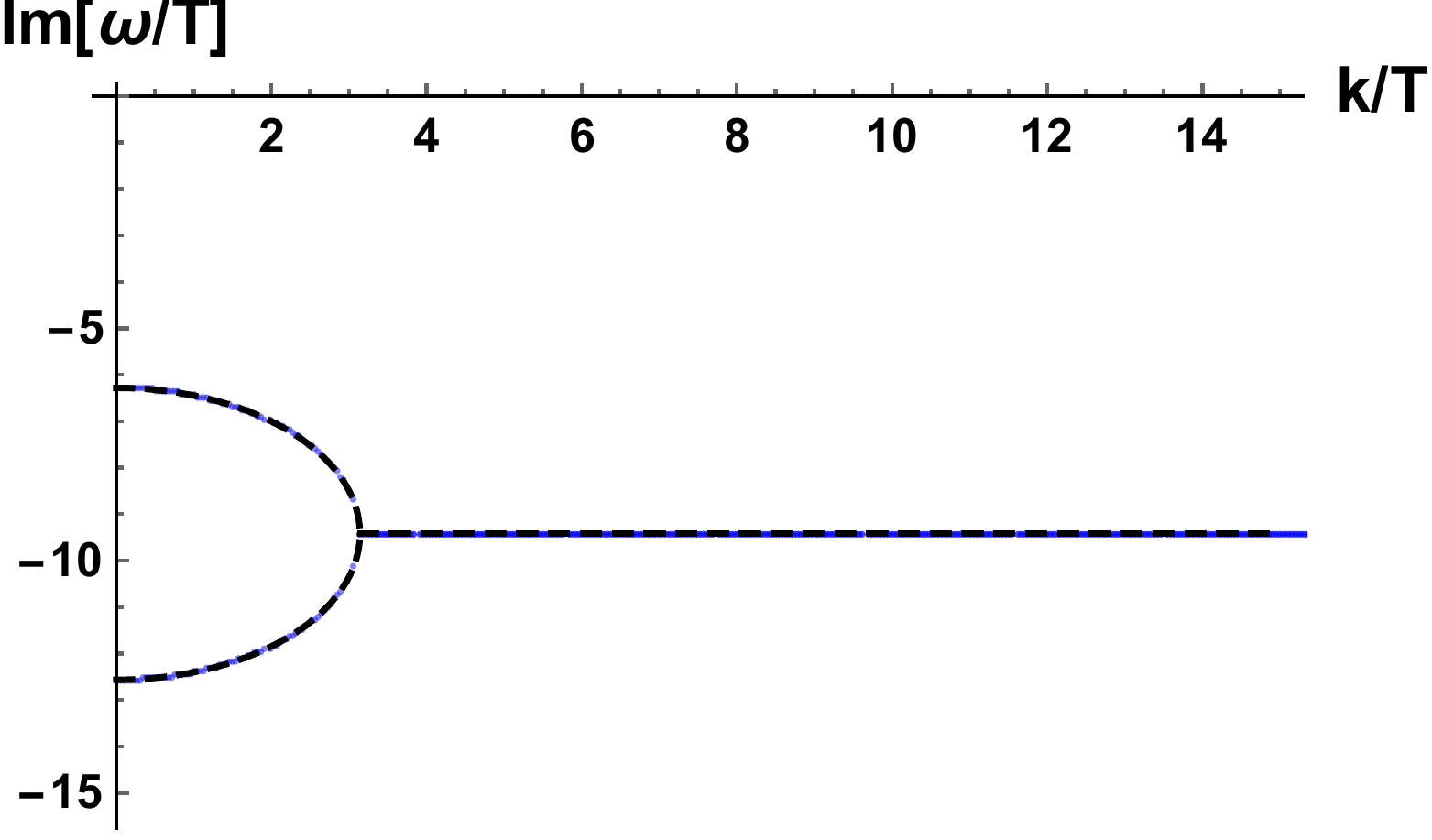}
\caption{A comparison between the analytic formula \eqref{selfF} obtained in \citep{Davison:2014lua} and the QNMs spectrum computed numerically at the self dual point $\epsilon=0$. The agreement is excellent.}
\label{fig4}
\end{figure}
For a specific value of $T/m \approx 0.159$ the energy density vanishes and the system enjoys an enhancement of the symmetries which allow to compute the correlators and their poles analytically as shown in \cite{Davison:2014lua}. For such a value the poles satisfy:
\begin{equation}
\omega\,=\,-\,\frac{3}{2}\,i\,\pm\,\sqrt{k^2\,-\,\frac{1}{4}}\label{selfF}
\end{equation}
which clearly shows the presence of a gap in momentum space $k_{gap}=1/2$. This results is confirmed by our numerical results in fig.\ref{fig4}. We will come back on the nature and origin of the k-gap in the next section.
\subsection{The k-gap and Maxwell interpolation}
\begin{figure}
\center
\includegraphics[width=7.5cm]{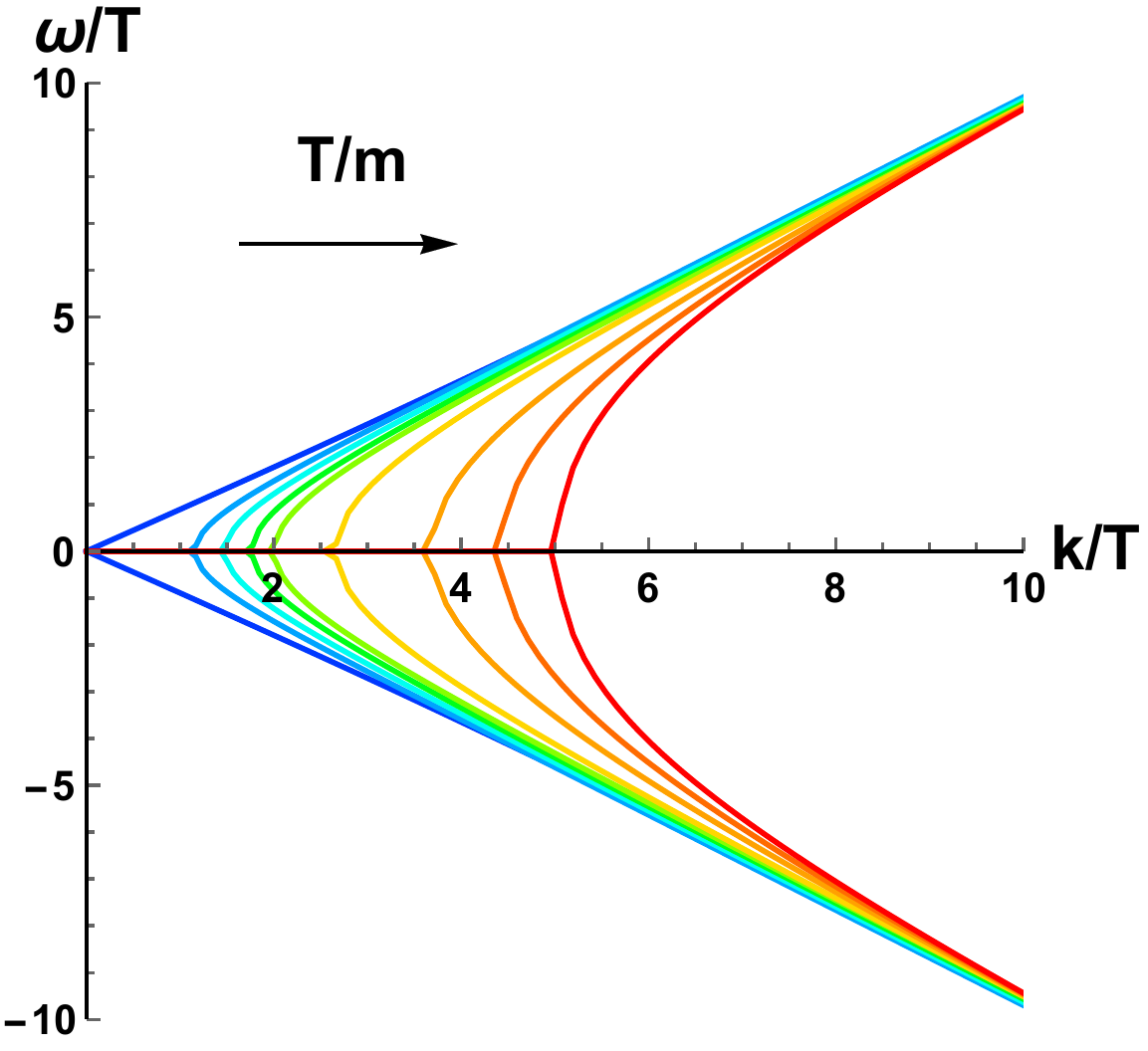}
\quad
\includegraphics[width=7.5cm]{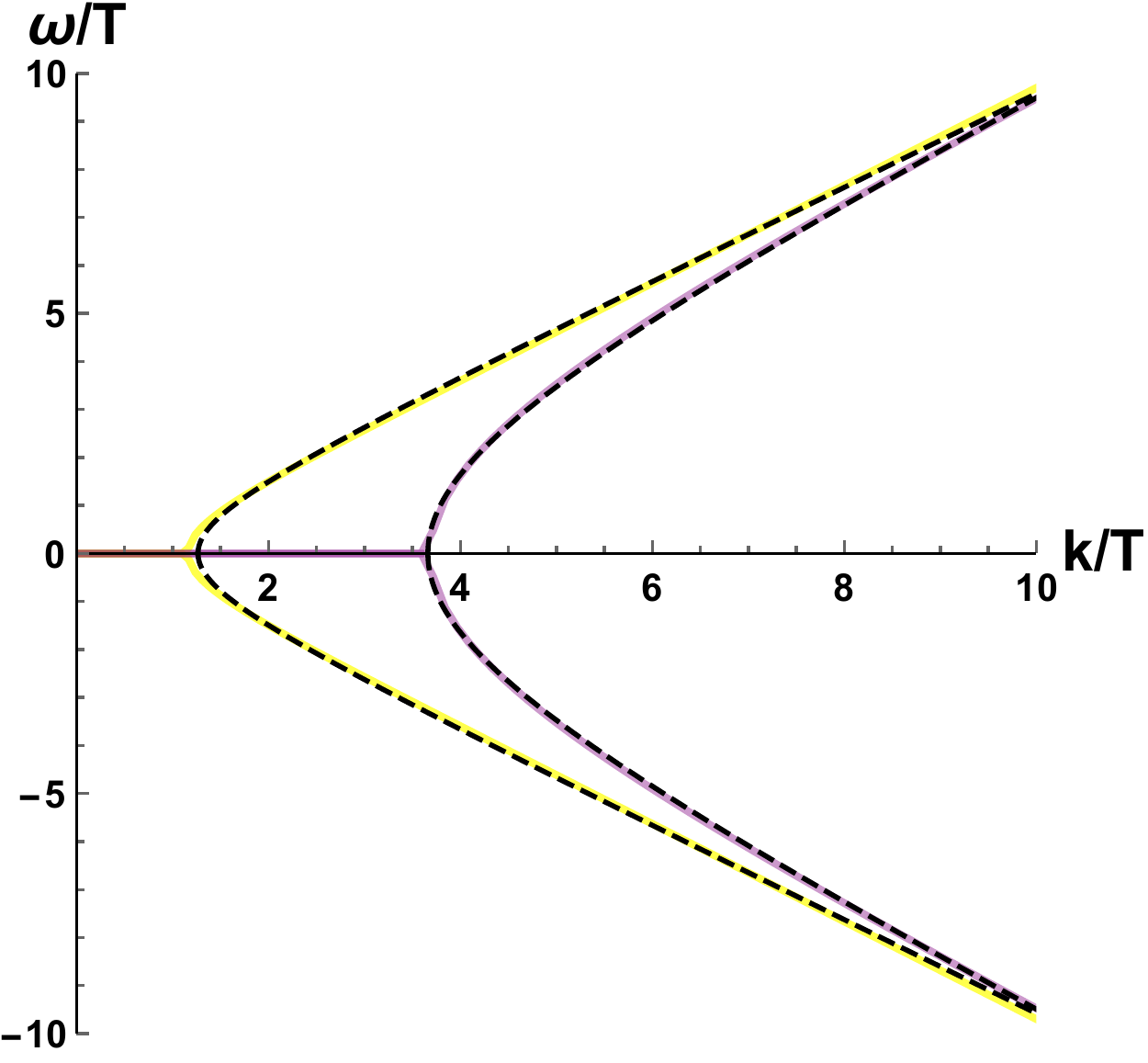}
\caption{\textbf{Left:} A collection of dispersion relations exhibiting the k-gap for increasing temperature. \textbf{Right:} Some benchmark results for the fit $\omega=\left(a^2\,k^2\,-\,b^2\right)^\nu$; $\nu$ is always compatible with the expected value $\nu=1/2$ and $a$ is compatible, up to numerical error, to the relativistic limit $a=1$ which is always reached at large momentum $k/T\gg 1$.}
\label{fig5}
\end{figure}

\begin{figure}
\centering
\includegraphics[width=7.25cm]{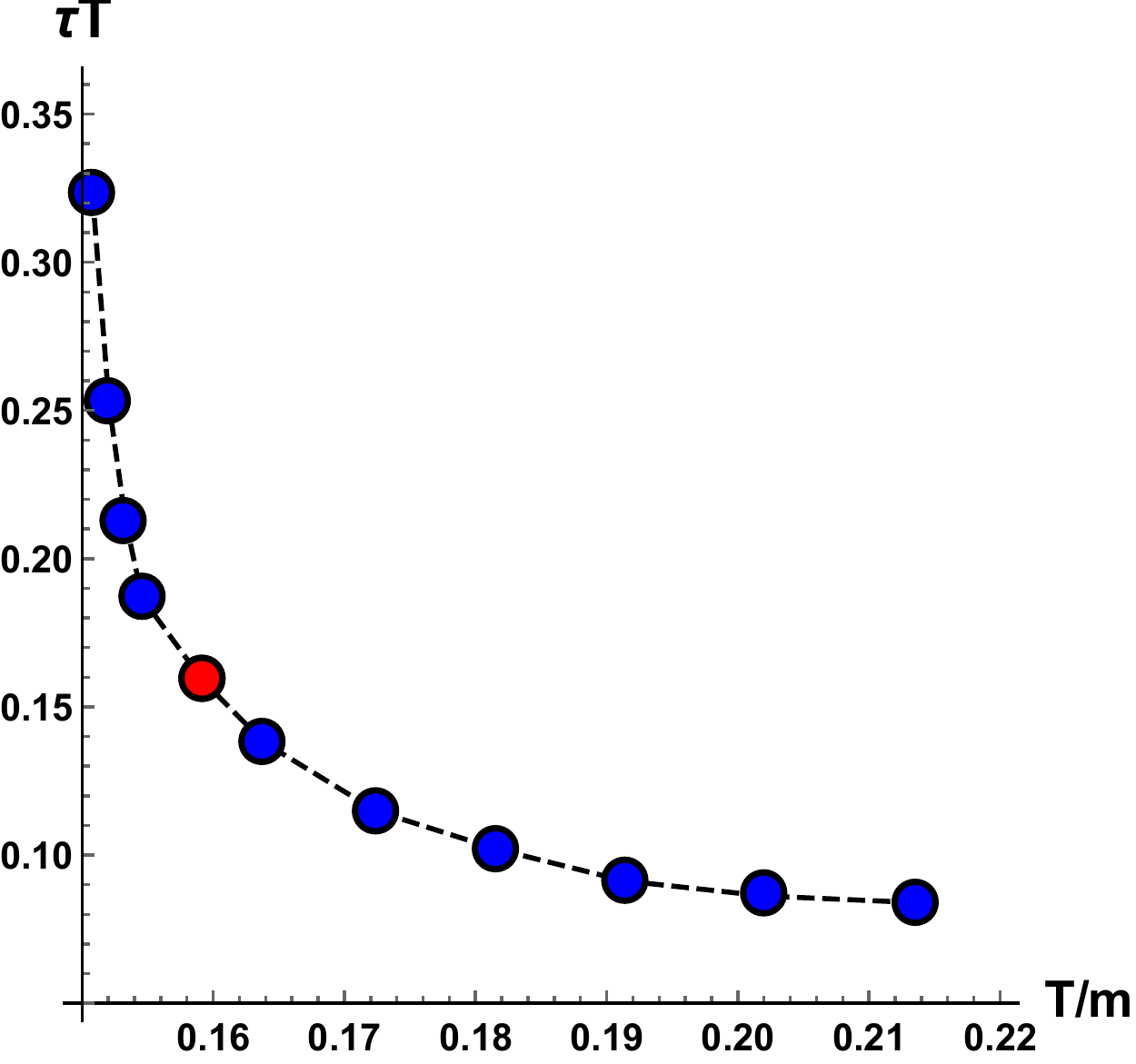}
\quad
\includegraphics[width=7.75cm]{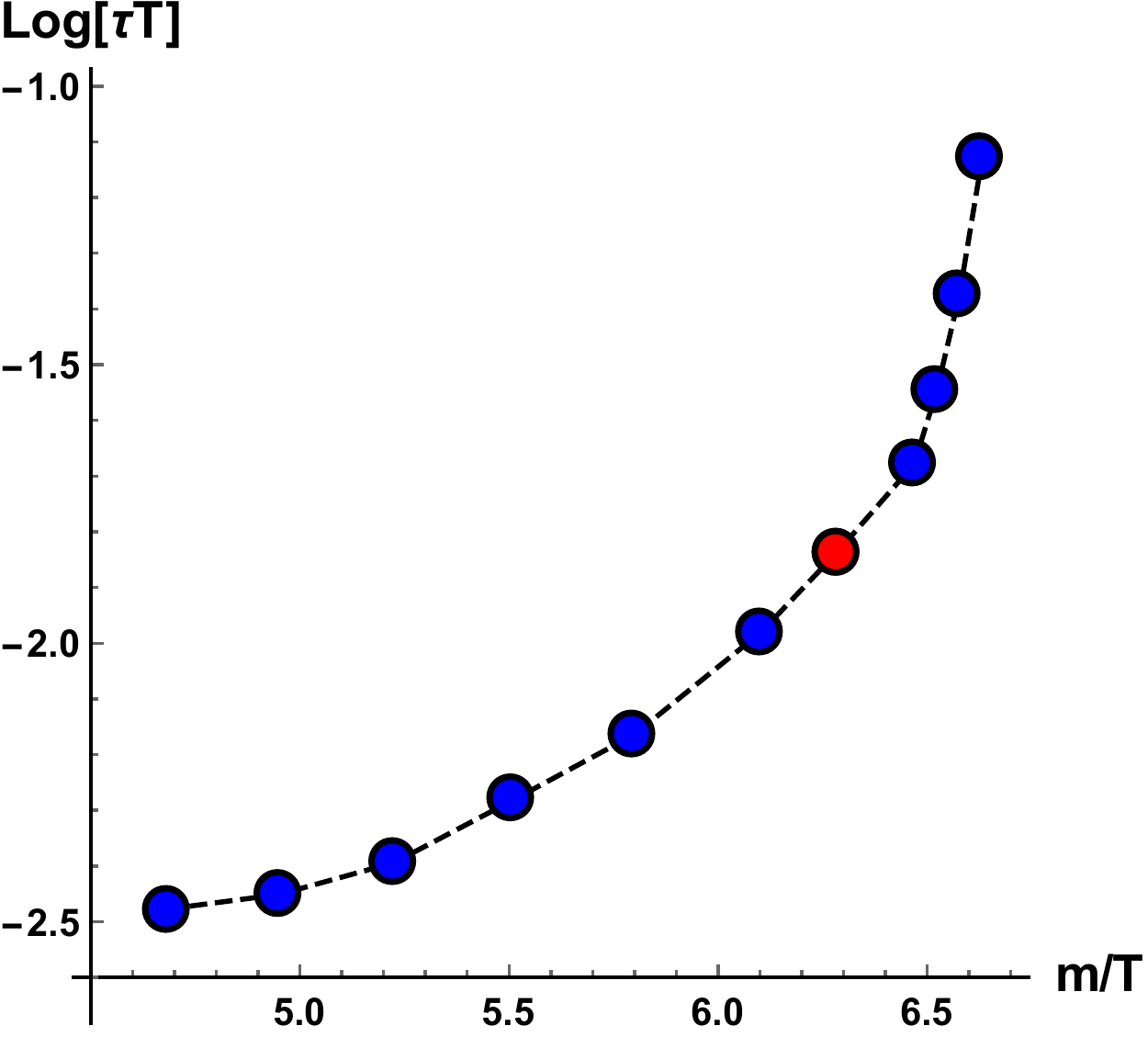}
\caption{The timescale $\tau$ governing the k-gap in function of temperature. The red dot is the analytic result at the ''self-dual'' temperature $T^\star$. The dashed lines are just joining together the numerical data. It is important to notice the similarities of this behaviour with the well-known Vogel-Tammann-Fulcher law for liquids.}
\label{fig6}
\end{figure}

Our main interest is the region in the parameter space where the energy density is positive and a gap in momentum space is present. In particular we emphasize that the presence of the $k$-gap is limited to the so-called \textit{coherent region}. In this region, the system lies in a coherent regime displaying a sharp Drude peak in the electric conductivity. In order to study this feature better, we collected more data for $T/m >0.159$. Some examples are shown in fig.\ref{fig5}. Our main purpose is to understand the origin of the emergence of the $k$-gap and make contact with the theories proposed in section \ref{prologue}.\\ 
The first comment, which simply comes from looking at fig.\ref{fig5} is that the k-gap moves towards higher momenta increasing the temperature of the system. This feature is exactly the same as that discussed in liquids earlier.\\
We now fit the curves to the functional form:
\begin{equation}
Re(\omega)\,=\,\sqrt{c^2\,k^2\,-\,\frac{1}{4\,\tau^2}}\label{fitfun}
\end{equation}
which has been proposed in \cite{PhysRevLett.118.215502} and discussed further in \cite{2016RPPh...79a6502T}. The agreement between the fits and the data is excellent (see fig.\ref{fig5}). From our data we observe that the asymptotic speed is always relativistic:
\begin{equation}
\omega\,=\,k\,\,\,\,\text{for}\,\,k\rightarrow \infty\,\,\quad \leftrightarrow\quad\,c\,=\,1
\end{equation}
Such a feature is intimately connected to the presence of a relativistic and conformal UV fixed point. In CM language, our setup lacks the existence of a UV cutoff which can be identified with the lattice spacing; as a consequence the behaviour of our system is not cut by any UV scale and it differs from the condensed matter systems.\\
From the numerical fit we can obtain the behaviour of the timescale $\tau$ in function of the temperature of the system, which is shown in fig.\ref{fig6}. The timescale $\tau$ decreases with temperature and it shows a behaviour consistent with real liquids (see \cite{2016RPPh...79a6502T}). Additionally we observe that, defining the critical ''self-dual'' temperature $T^\star$, there is a clear change in the behaviour of $\tau$ below and above such a temperature scale as it is clear from the LogPlot in fig.\ref{fig6}.
\subsection{The relaxation time}
It is now interesting to consider the following questions. First, which physical quantities set the timescale $\tau$? Does such timescale coincide with the Maxwell relaxation time? Is the relaxation time derivable in terms of the hydrodynamic diffusion constant as derived in \eqref{fff}?

\begin{figure}
\centering
\includegraphics[width=7.5cm]{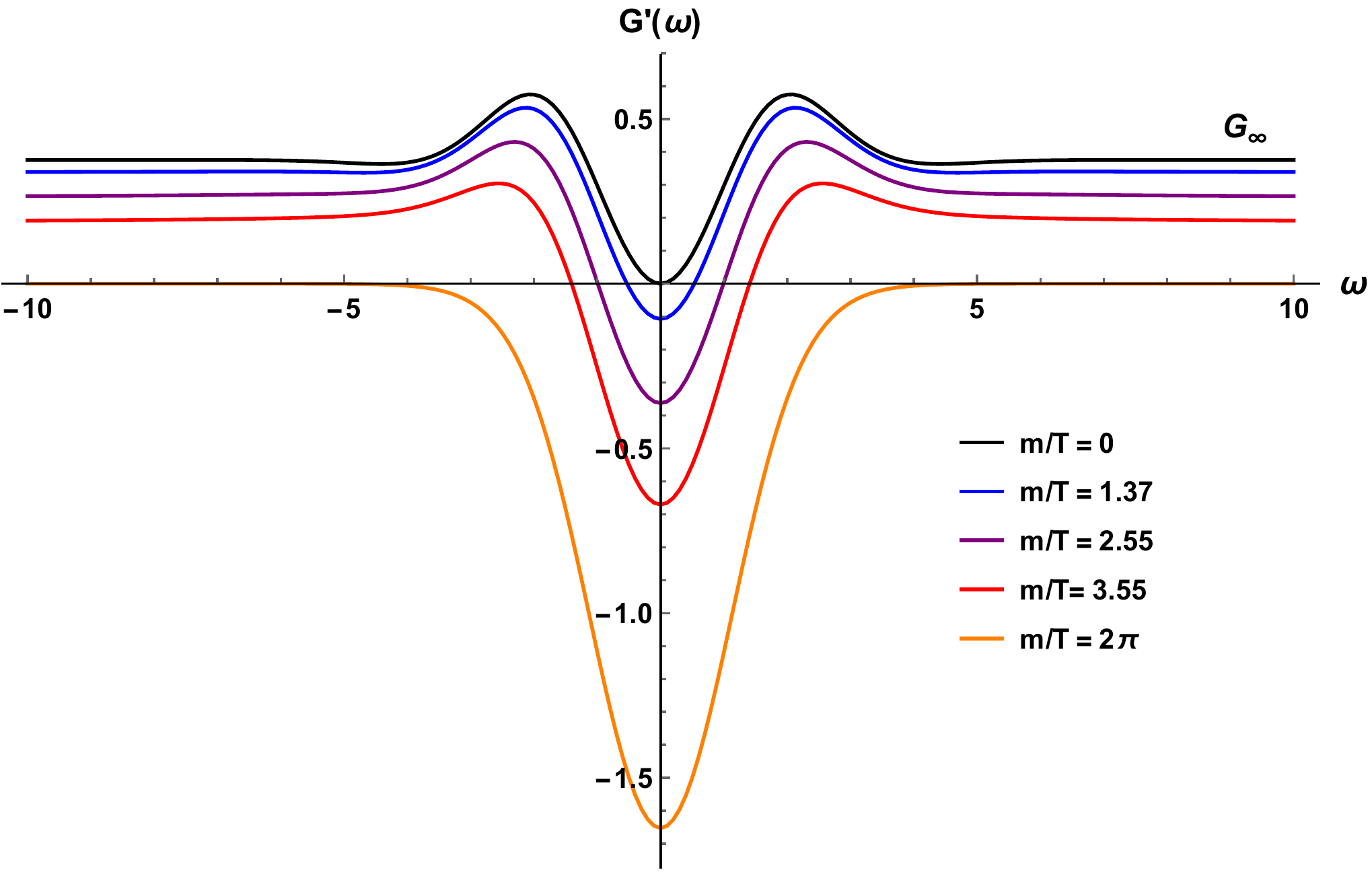}
\quad
\includegraphics[width=7.5cm]{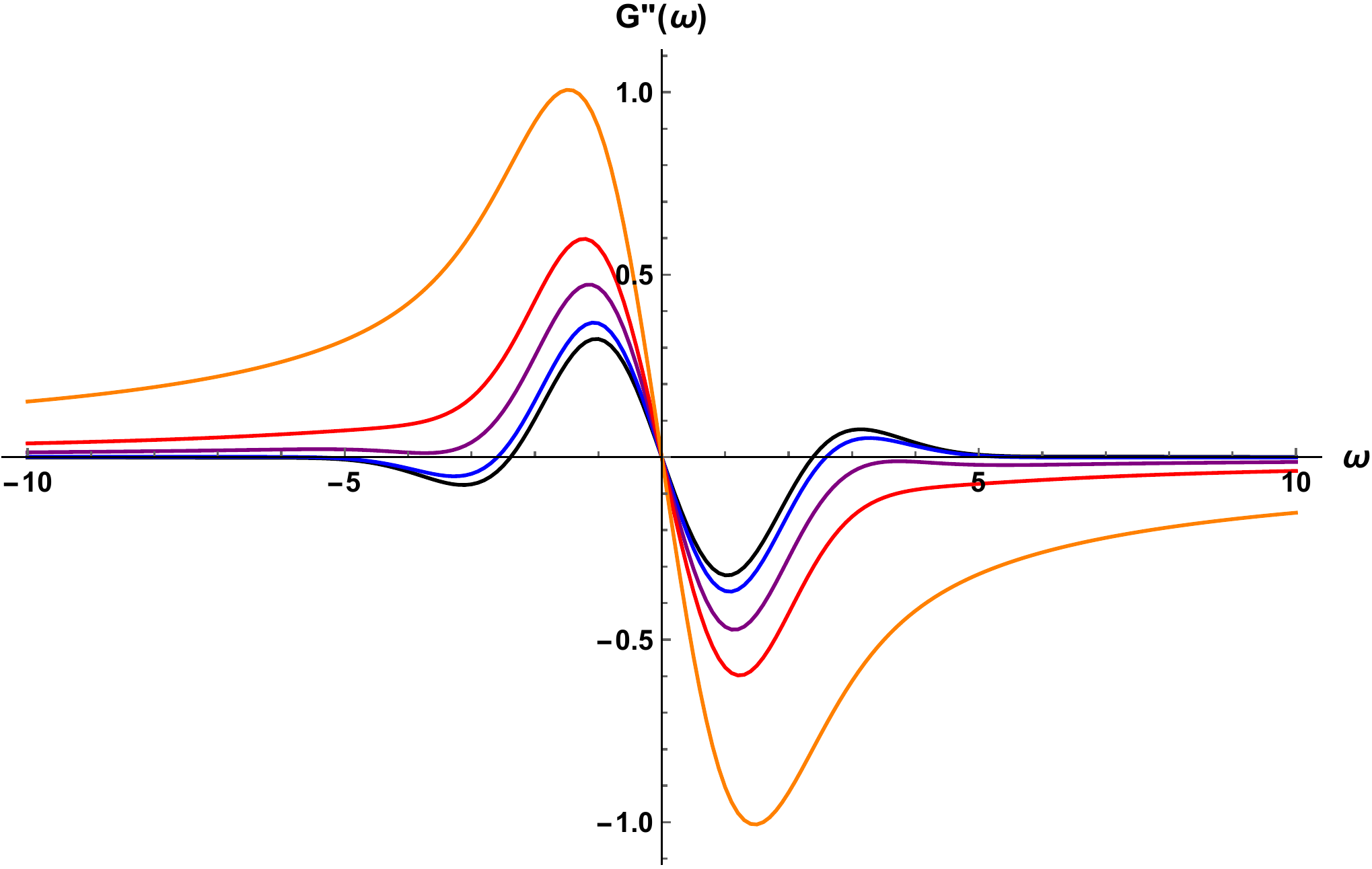}
\caption{The stress tensor Green function \eqref{corr} in function of the frequency for various values of $m/T$. \textbf{Left: }The real part $G'(\omega)$. \textbf{Right: }The imaginary part $G''(\omega)$.}
\label{fignew}
\end{figure}

\begin{figure}
\centering
\includegraphics[width=7.5cm]{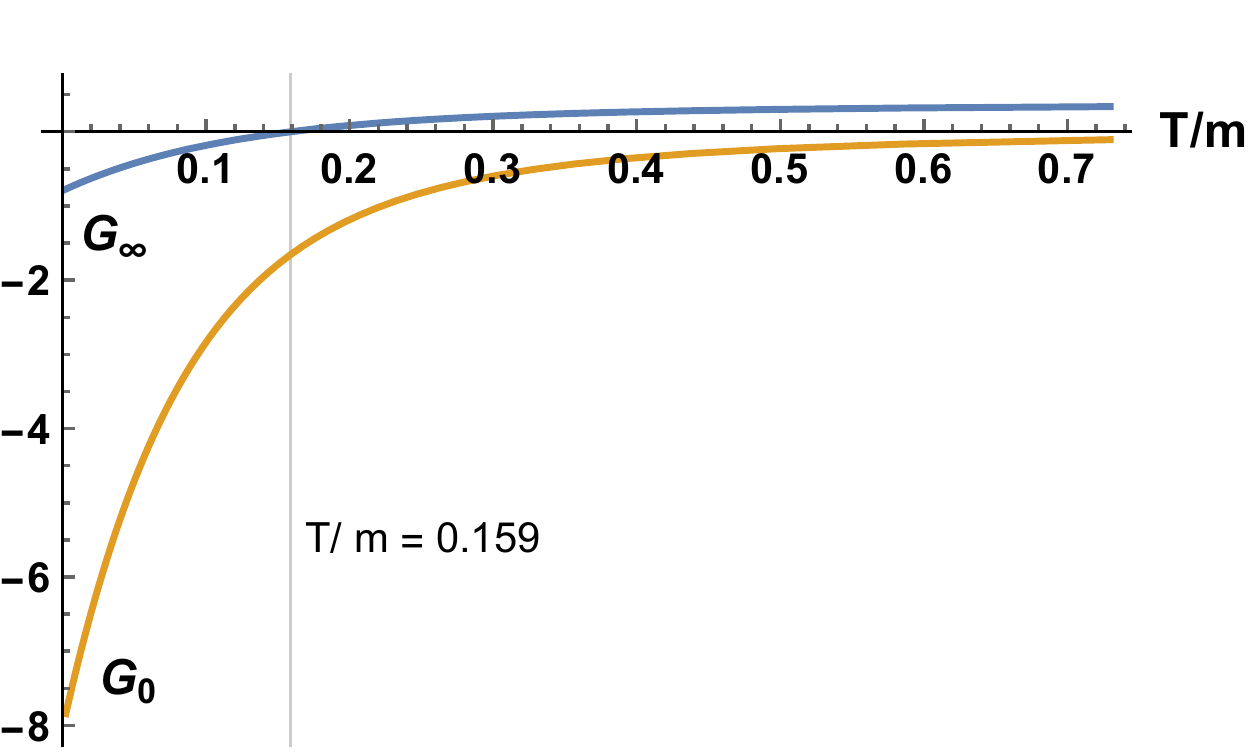}
\quad
\includegraphics[width=7.5cm]{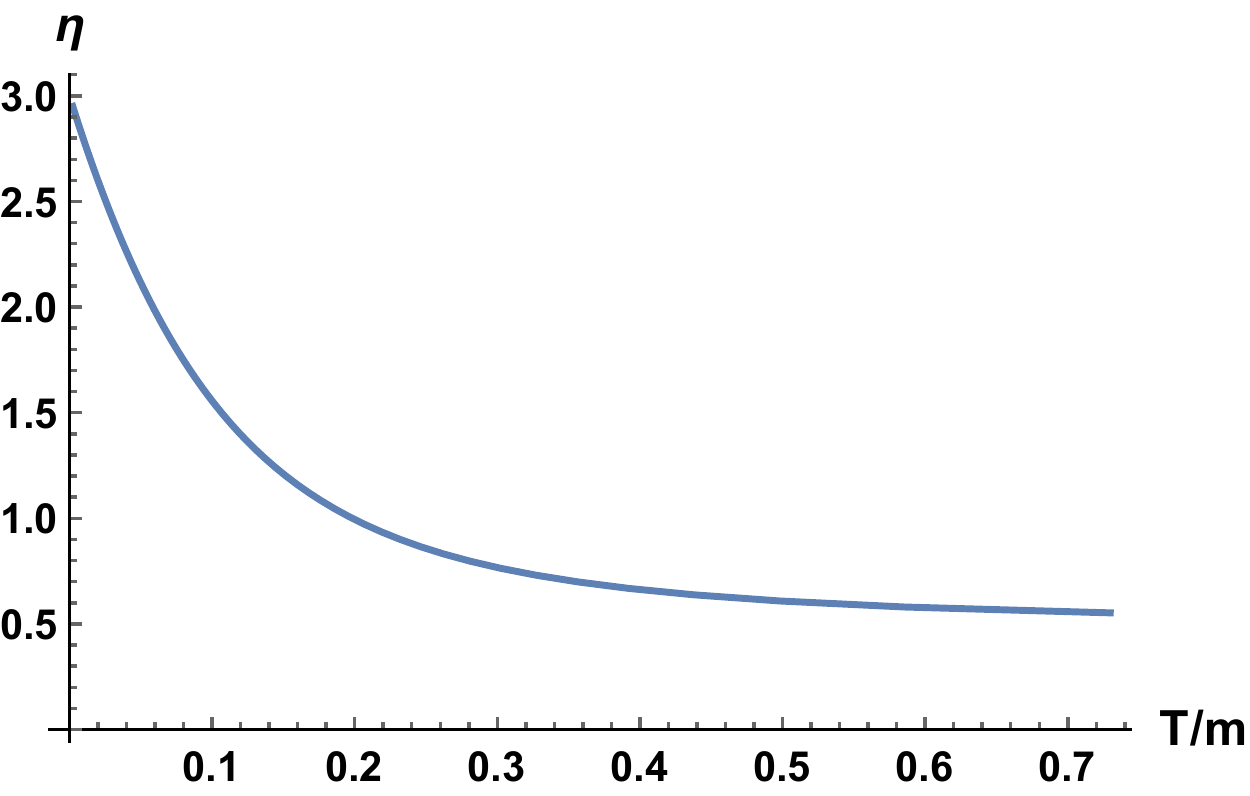}
\caption{\textbf{Left: } The static and infinite frequency elastic moduli $G_0$ and $G_{\infty}$ in function of $T/m$. \textbf{Right: }The shear viscosity in function of $T/m$. The asymptotic value at $T/m \rightarrow \infty$ is exactly $s/4\pi$.}
\label{fignewb}
\end{figure}

\begin{figure}
\centering
\includegraphics[width=8cm]{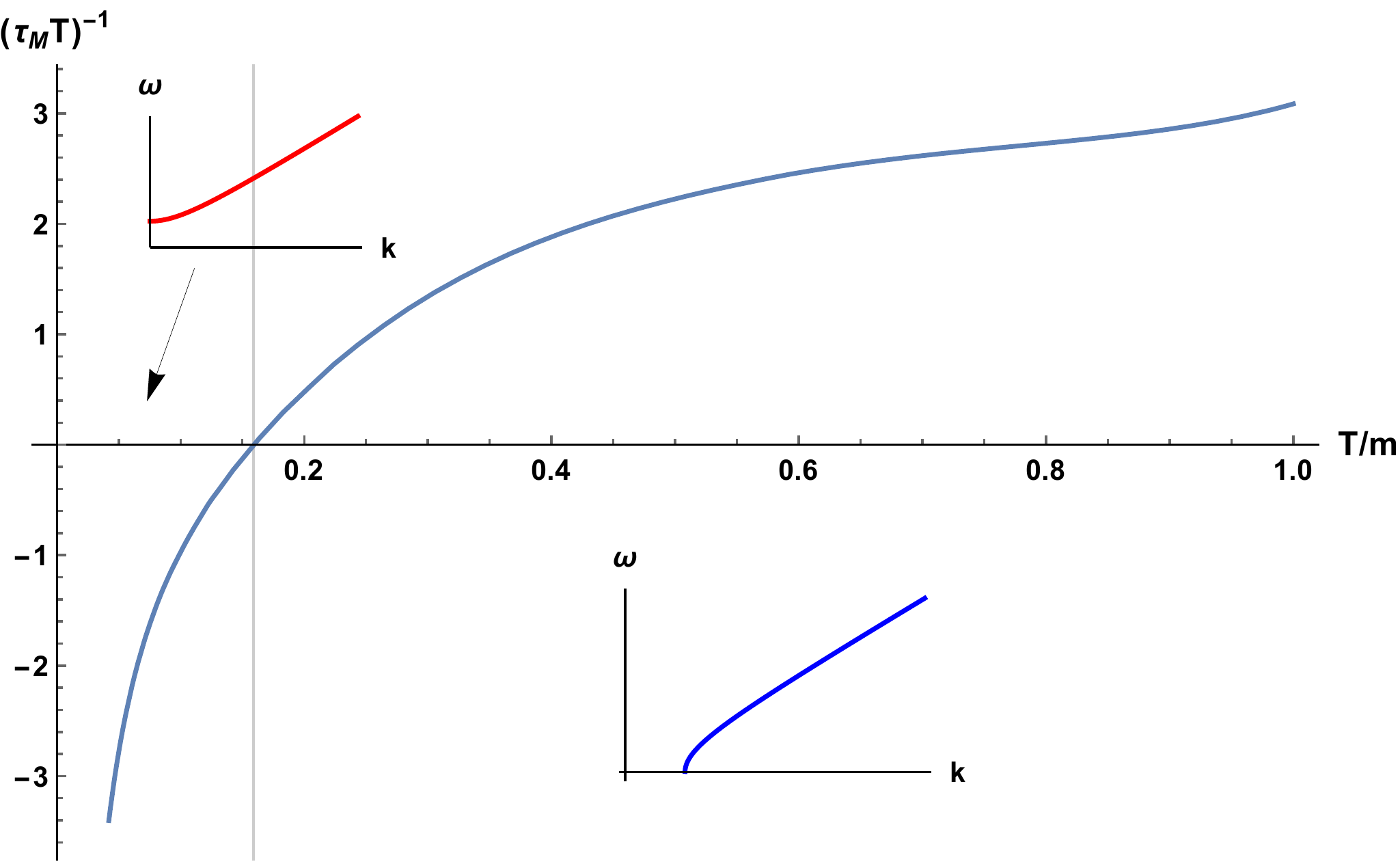}
\quad
\includegraphics[width=7cm]{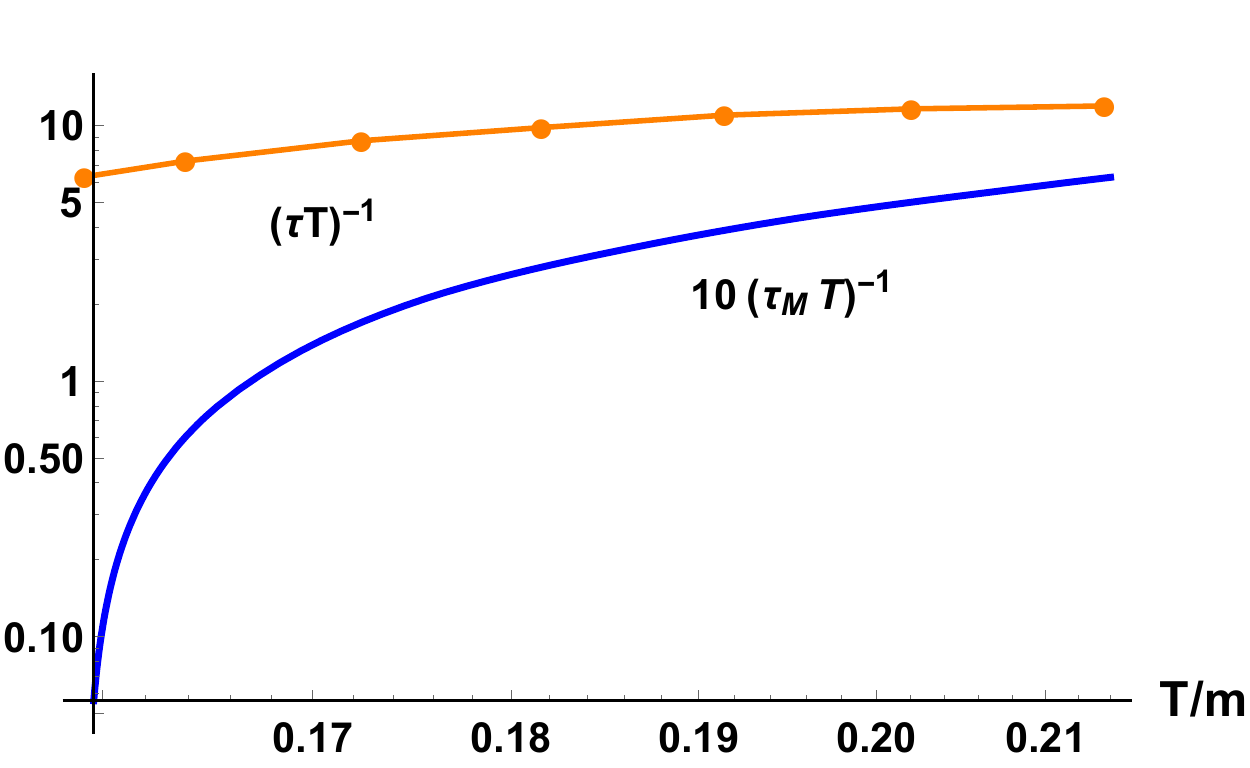}
\caption{\textbf{Left: }The Maxwell relaxation time $\tau_M=\eta/G_{\infty}$ in function of the dimensionless temperature $T/m$. Roughly the change of sign of the Maxwell timescale appears when the dispersion relation of the shear modes switch from massive particle type (red inset) to k-gap type (blue inset). \textbf{Right: }A comparison between the Maxwell relaxation time $\tau_M$ and the numerical data.}
\label{fignewc}
\end{figure}

In order to answer the questions above within this model we have first to compute the Maxwell relaxation time:
\begin{equation}
\tau_M\,\equiv\,\eta/G_{\infty}
\end{equation}
In order to do that we need to compute the shear correlator of the system at finite frequency. We can simply follow the computations of \cite{Alberte:2016xja}. In particular we have to solve the equation for the fluctuation $\delta g_{xy}=e^{-i \omega t}\,h(u)$ which reads in Eddington-Filkenstein coordinates:
\begin{equation}
h''\,+\,h'\,\left(-\frac{2}{u}\,+\,\frac{2\,i\,\omega}{f}\,+\,\frac{f'}{f}\right)\,+h\,\left(-\frac{2\,i\,\omega}{u\,f}\,-\,\frac{2\,m^2}{f}\right)\,=\,0\,.
\end{equation}
where $f(u)$ is the metric emblackening factor and we use standard notations for the derivative with respect to the radial coordinates $u$.\\ Now given that the UV asymptotics of the $h(u)$ field are:
\begin{equation}
h(u)\,=\,h_0(\omega)\,+\,h_3(\omega)\,u^3\,+\,\dots
\end{equation}
close to the AdS boundary $u=0$; the Green function for the shear stress tensor operator is simply given by:
\begin{equation}
G^R_{T_{xy}T_{xy}}(\omega)\,=\,\frac{3}{2}\,\frac{h_3(\omega)}{h_0(\omega)} \label{corr}
\end{equation}
where the index $R$ reminds the reader that we are considering the retarded correlator.\\
Now we can directly define the viscosity of the system and the infinite frequency elastic modulus as:
\begin{align}
&\eta\,=\,-\,\lim_{\omega \rightarrow 0}\,\frac{1}{\omega}\,G''(\omega)\,,\,\quad\, G_{\infty}\,=\,\lim_{\omega \rightarrow \infty}\,G'(\omega)
\end{align}
where we have used the reduced notation:
\begin{equation}
G^R_{T_{xy}T_{xy}}(\omega)\,=\,G'(\omega)\,+\,i\,G''(\omega)
\end{equation}
splitting the real and imaginary parts of the Green function.\\
The results for the shear correlator \eqref{corr} are shown in fig.\ref{fignew}.
As expected the real part of the correlator asymptotes a constant value at infinite frequency which indeed we identify with $G_{\infty}$. The values of the static elastic modulus, the infinite frequency elastic modulus and the shear viscosity are shown in fig.\ref{fignewb}.
Several interesting properties appear. First, the static and infinite frequency moduli clearly do not match. The static elastic modulus is always negative, while the infinite frequency one is positive for large enough temperatures. More precisely we can prove that:
\begin{equation}
G_\infty\,=\,\frac{3}{8}\,\epsilon\,,
\end{equation}
where $\epsilon$ is the energy density of the system. Therefore the infinite frequency modulus becomes positive around $T/m \sim 0.159$ which is indeed the self-dual point where $\epsilon=0$. Interestingly enough this is approximately also the value at which the dispersion relation of the shear modes passes from the k-gap type to the massive particle type. See for example the previous fig.\ref{3}. This is an important point because it implies that the Maxwell relaxation time is well-defined only for temperatures $T/m > 0.159$ which is roughly the temperature at which the k-gap appears. The final result for the Maxwell relaxation time is displayed in fig.\ref{fignewc}.
Comparing the Maxwell relaxation time $\tau_M$ with the relaxation time $\tau$ extracted from the numerical data we can claim that the Maxwell result overestimates the correct value by more than one order of magnitude. Moreover the Maxwell relaxation time becomes negative at $T/m \sim 0.159$ while the k-gap closes ($\tau=0$) at a slightly lower temperature. Nevertheless it is interesting to notice that in the region where both the relaxation times are well defined the qualitative behaviour is very similar. As a matter of fact both the timescales decrease increasing the temperature (see fig.\ref{fignewc}).\\
As a short summary of this analysis we demonstrated that the relaxation time in our holographic model does not coincide with the Maxwell time extracted from holography; nevertheless the qualitative temperature dependence is rather similar.

\begin{figure}
\centering
\includegraphics[width=8cm]{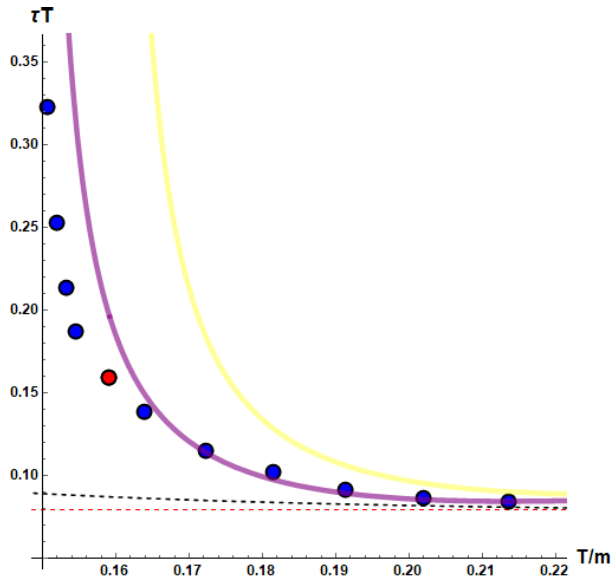}
\caption{A comparison between the relaxation time $\tau$ numerically computed from the QNMs (filled bullets) and several hydrodynamic quantities. (I) the diffusion constant $D$ (yellow), (II) the formula for the generalized liquid relaxation time $\tau$ \eqref{new} (purple) (III) the approximated formula for the diffusion constant \eqref{anal} (dashed). Notice that for $\Gamma \tau \ll 1$ the corrections from $\tau_{rel}$ are small and $\tau \approx D/c^2$. In general we found a good agreement between the data and the theory. The red bullet indicates the self-dual point.}
\label{check1}
\end{figure}

Discarded the hypothesis that the relaxation time is controlled by the Maxwell timescale $\tau_M$ it is still interesting to understand if the relaxation time can be related to the hydrodynamic diffusion constant as introduced in \eqref{fff}.
In order to answer this question within this model we have to take into account that hereby we have a finite relaxation time for the momentum operator $\tau_{rel}\neq 0$ and therefore the simple quasihydrodinamic picture is expected to work just at high temperature (where momentum relaxation is negligible) as shown in fig.\ref{check1}. Due to the aforementioned complication we have to modify our setup and consider that our total relaxation time appearing in \eqref{fitfun} contains also a contribution from the momentum relaxation time $\tau_{rel}$. In particular, in the limit of  $\Gamma/T \ll 1, \tau T \gg 1$ the dispersion relation of the two first modes can be described by:
\begin{equation}
\omega_{\pm}\,=\,-\,i\,\Gamma\,-\,\frac{i}{2\,\tau}\,\pm\,\sqrt{c^2\,k^2\,-\,\frac{1}{4\,\tau^2}\,+\,\dots}
\end{equation}
which once expanded in the low momentum limit $k/T \ll 1$ gives rise to the two modes:
\begin{equation}
\omega\,=\,-\,i\,\Gamma\,-\,i\,D\,k^2+\dots,\quad \omega\,=\,-\,i\,\Gamma\,-\,\frac{i}{\tau}\,+\,+\,i\,D\,k^2\,\quad D\,\equiv\,c^2\,\tau
\end{equation}
which is exactly what our numerical data show (see for example fig.\ref{fig4}). It is important here to underline an additional difficulty of our setup. The above description is only a good one when $\Gamma \ll T, \tau \gg 1/T$. The condition of having the first ''Drude'' pole with a small damping indeed implies $\Gamma/T \ll 1$; at the same time we have to demand that the second pole is also not overdamped such that the hydrodynamic description can be written down just in terms of these two modes. Putting together the two constraints, one would naively expect a good agreement between the formulas and the numerics only in an intermediate regime of temperatures. At very large values of $m/T$, the momentum relaxation becomes important removing the basis for the approximation of having a \textbf{underdamped} pole $\omega=-i \Gamma$. In the opposite limit at large temperature, the other pole will acquire a very large imaginary part because $\tau^{-1}$ will grow with temperature and therefore it will disappear from the hydrodynamic spectrum. Let us notice that the limit $\tau T \gg 1$ implies also that $k_{gap}/T \ll 1$, namely that the $k-$ gap lies inside the hydrodynamic window. Our numerical data confirm that, even beyond that limit where our ''hydrodynamic'' formulas do not work anymore, a well defined $k-$gap appears. In that sense, we can think about the $k-$ gap phenomenon as a beyond-hydrodynamic feature. We notice that, in the regime of validity of our theoretical framework, the relaxation time is well approximated by :
\begin{equation}
\tau\,=\,\frac{D}{c^2}\,=\,D
\end{equation}
since $c=1$. Decreasing the temperature, outside the range of validity of our formulas because of the strong momentum relaxation, we find that the relaxation time is better fitted by the expression\footnote{Let us remark that this expression cannot be derived in the context of quasyhydrodynamics.}:
\begin{equation}
\tau\,=\,\frac{D}{c^2\,+\,D\,\Gamma}\,<\,\frac{D}{c^2}\label{new}
\end{equation}
which can be guessed using a, non justified, inverse Matthiessen, and assuming $\tau_{tot}^{-1}=\tau^{-1}+\Gamma=c^2/D$.\\
Notice that in the limit $\Gamma/T \ll 1$ the corrections coming from momentum dissipation are negligible.
As a summary of this second attempt we emphasize that in the regime of validity of our approximation, i.e. at large temperature, the relaxation time is indeed given by the hydrodynamic diffusion constant as predicted by formula \eqref{fff}.\\

\begin{figure}
\centering
\includegraphics[width=7cm]{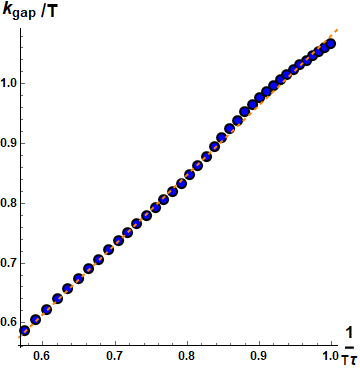}
\caption{The k-gap in function of the inverse relaxation time $\tau$ \eqref{new}. The straight dashed line is a guide for the eye}
\label{un}
\end{figure}

Finally, we address the behaviour of the k-gap as a function of the inverse of the relaxation time $1/\tau$ where we fix $c=1$. The result is shown in fig.\ref{un}. The curve shows a linear behaviour, in agreement with what Eq. (\ref{kgap}) predicts and what was found in liquids and supercritical fluids \cite{PhysRevLett.118.215502}. This is another important result highlighting similarities between liquids and holographic models.
\label{hol2}
\section{Discussion}\label{disc}
In this work we performed a detailed study of the shear collective modes beyond the hydrodynamic limit in two simple holographic bottom-up models exhibiting viscoelastic features. Our main result is that both models show the presence of k-gap in the transverse spectrum whose temperature dependence is in perfect agreement with what is seen in condensed matter systems, liquids. Since the gap and its properties in liquids are derived from the Maxwell interpolation, our results imply that relativistic, and strongly coupled, systems and systems with slow momentum relaxation also obey Maxwell interpolation. Nevertheless the relaxation time of the holographic models considered is not given by the Maxwell relation timescale $\tau_M$.  In the hydrodynamic limit, or more precisely when $k_{gap}/T \ll 1$ and for the second model $\Gamma/T \ll 1$, the corresponding expression for the liquid relaxation time reads:
\begin{equation}
\boxed{\tau\,=\,\frac{D}{v^2}}
\end{equation}
where $D$ is the hydrodynamic diffusion constant extracted by the dispersion relation of the (pseudo)-diffusive mode at $k/T \ll 1$ and $v$ the asymptotic transverse sound speed at large momentum. At low temperature, where the momentum relaxation rate is not small anymore, we found a good agreement with the empirical formula:
$\tau\,=\,D/(c^2\,-\,D\,\Gamma)$
where $\Gamma=\tau_{rel}^{-1}$ is the momentum relaxation rate. Interestingly, at zero momentum dissipation $\Gamma=0$, substituting into the previous formula the expression for the diffusion constant in a relativistic fluid $D=\eta/\chi_{PP}$ and the formula for the transverse sound speed $v^2=G/\chi_{PP}$ we immediately re-derive the Maxwell interpolation result $\tau=\eta/G$.\\[0.2cm]
We summarize our findings in the following:
\begin{itemize}
\item The thermodynamic properties of the considered models are in agreement with the expectations from realistic viscoelastic materials \cite{RevModPhys.78.953,ped}. In particular the temperature dependence of the heat capacity displays a crossover between the Debye model $c_v\sim T^d$ at large temperature and a new emergent linear behaviour at small temperature $c_v \sim T$.  Our results indicate that this effect can be associated to the onset of the incoherent-coherent transition \cite{Davison:2014lua}. From the geometrical point of view this is just a manifestation of the UV AdS$_4$ and the IR AdS$_2\times R^2$ fixed points.
\item Both holographic models show the existence of a k-gap in the transverse spectrum of excitations (beyond the hydrodynamic limit) as is seen in liquids and supercritical fluids \cite{PhysRevLett.118.215502}. The $k$-gap in the holographic models is strikingly similar to the gap in liquids: it (a) increases with temperature and (b) linearly increases with the inverse of relaxation time.
\item In both the holographic models the dispersion relation of the propagating shear waves is in agreement with the Frenkel proposal\
\begin{equation}
Re(\omega)=\sqrt{v^2 k^2-\frac{1}{4 \tau^2}}\label{last}
\end{equation}
which can be phenomenologically derived from the Navier Stokes equation using the substitution \eqref{sub}.
\item Despite the previous agreement, the relaxation time scale in eq.\eqref{last} is not given by the Maxwell prediction $\tau_M=\eta/G_{\infty}$. The only similarity between the two timescales is their qualitative temperature dependence, indeed both decreasing with temperature. Additionally such a dependence is similar to what is widely observed in liquids \cite{2016RPPh...79a6502T} (see fig.\ref{fig6}).
\end{itemize}

Our results strongly suggest that Maxwell interpolation has a much wider scope of application than envisaged originally and discussed until now despite the identification of the relaxation timescale has a more complicated formula with respect to the one proposed by Maxwell. Namely, the ability of the system to support both hydrodynamic viscous and solid-like elastic response and, importantly, the additivity of both responses in the equation of motion (see \ref{a1}) are the two properties which we believe are applicable in other systems. Importantly, this approach applies to strongly-coupled systems where neither elastic nor viscous components are small and where perturbation theory and other existing theories do not apply. For this reason, we believe that the approaches based on Maxwell interpolation will be fruitful in addressing strongly-coupled field theories which describe a number of important physical systems and phenomena. A more detailed discussion of this point will be given in a companion paper \cite{PRL}.\\[0.2cm]
As an interesting observation we should underline the similarity of the picture described with the phenomenological Israel-Stewart model \cite{ISRAEL1979341} and the related higher derivative holographic models \cite{Grozdanov:2016vgg,Grozdanov:2016fkt}. In other words the presence of the k-gap is somehow expected and needed to regulate the well-known problems of acausality produced by the diffusive hydrodynamic pole. See for example \cite{Romatschke:2009im}.\\[0.2cm]

Several questions and future directions remain to be investigated:
\begin{itemize}
\item The k-gap phenomenon appears to be very general within the holographic constructions. Its presence has already been observed in several contexts: in the dispersion relation of chiral magnetic waves within the anomalous magneto response \cite{Jimenez-Alba:2014iia}, in plasmas with finite magnetic field \cite{Grozdanov:2016tdf,Grozdanov:2017kyl}, in the study of electromagnetism using global symmetries \cite{Hofman:2017vwr}, in the QNMs spectrum of P-wave superfluids \cite{Arias:2014msa} and in holographic fermionic spectral functions\footnote{\textit{Private communication with Sang-Jin Sin and  
Yunseok Seo.}}. It would be interesting to better understand the connection between the  physics of the k-gap in liquids, the emergence of propagating shear waves and the features of Alfven waves and emergent photons. Finally, is there a common physical origin to the $k$-gap in many physical systems and effects mentioned above? Is it related to a finite range of propagation of an excitation in a medium where this excitation is not an eigenstate or, in other words, to dissipation? In this simple yet general picture, an existence of a finite relaxation time $\tau$ in the system implies a finite range of wave propagation equal to $c \tau$. Then, the $k$-gap $k_g=1/(c \tau)$ emerges simply because the allowed wavelengths can not be longer than the wave propagation range?
\item It is very interesting that we can define the $k$-gap just using hydrodynamic observables such as the diffusion constant. \textit{Can the $k$-gap a new and valuable method to measure the diffusion constant in relativistic fluids?}
\item The physics of the k-gap is hiddenly connected with the so-called \textit{zero sound} in D-branes setups which follow from a DBI-type action \cite{Karch:2008fa,HoyosBadajoz:2010kd}. Moreover, our equation \eqref{mm} already appeared in this context in \cite{Chen:2017dsy} and it might also be relevant for the hydrodynamic description of condensed matter systems like graphene \cite{2016PhRvB..93x5153L}. Is there a unified picture of the k-gap phenomenon and the physics behind it?
\item Can we obtain an independent derivation of the liquid relaxation time formula \eqref{new}? Can large-D methods \cite{Andrade:2015hpa} be useful in this direction? How generic is the presence of a k-gap ? Can we re-derive our formula from field theory methods and appropriately deforming the hydrodynamic equations? We expect that our holographic results would agree with that analysis.
\item The similarities between the thermodynamic and the transverse spectrum features of our holographic models and those of realistic liquids are suggestive. It is interesting to study this further and the relationships with the instantaneous normal modes present in real liquids \cite{PhysRevLett.83.108}.
\end{itemize}

Finally, our discussion of the differences between liquids and gases has shown that important and unexpected liquid properties can be understood on the basis of Maxwell-Frenkel picture combining both hydrodynamic and solid-like elastic responses from the outset. This points to a new avenue to be investigated in regard to the cherished link between the holographic methods and experimental low energy results.

\section*{Acknowledgments}
We thank Tomas Andrade, Panagiotis Betzios, Alex Buchel, Richard Davison, Carlos Hoyos, Karl Landsteiner, Daniele Musso, Laurence Noirez, Napat Poovuttikul,Paul Romatschke and Alessio Zaccone,  for useful discussions and comments about this work and the topics considered. We are particularly grateful to Martin Ammon and Amadeo Jimenez for help, support and discussions and for sharing with us the numerical codes used in \cite{Alberte:2017cch}. We thank Martin Ammon, Tomas Andrade, Saso Grozdanov, Oriol Pujolas, Napat Poovuttikul and Alessio Zaccone for reading a preliminary version of this draft and provide useful comments and corrections.
We would like to thank Blaise Gouteraux, Ignacio Salazar Landea, Yunseok Seo, Sang-Jin Sin and Ben Withers for interesting discussions after the presentation of this work at Gauge-Gravity 2018. This research utilised Queen Mary's Apocrita HPC facility, supported by QMUL Research-IT.
KT is grateful to the Royal Society and EPSRC for support.
MB is supported in part by the Advanced ERC grant SM-grav, No 669288.
MB would like to thank Enartia Headquarters for the hospitality during the completion of this work. MB would like to thank Marianna Siouti for the unconditional support.

\appendix

\bibliographystyle{JHEP}
\bibliography{kgap2}
\end{document}